\documentclass{jfm}

\usepackage{graphicx}
\usepackage{newtxtext}
\usepackage{newtxmath}
\usepackage{natbib}
\usepackage{hyperref}

\newcommand{\RomanNumeralCaps}[1]
\linenumbers

\usepackage{float}
\usepackage{lipsum}  
\usepackage{siunitx}  
\usepackage{comment}
\usepackage{diagbox}
\usepackage{booktabs}
\usepackage{gensymb}
\hypersetup{
    colorlinks = true,
    urlcolor   = blue,
    citecolor  = black,
}
\usepackage{orcidlink}

\newcommand{\pder}[2]{\frac{\partial #1}{\partial #2}} 

\title{Shape modes and jet formation on ultrasound-driven wall-attached bubbles}

\author{M. Cattaneo\aff{1}
  \corresp{\email{mcattaneo@ethz.ch}},
  L. Presse\aff{1},
  G. Shakya\aff{1},
  T. Renggli\aff{1},
  B. Lukić\aff{2},
  A. Prasanna\aff{1},
  D. W. Meyer\aff{1},
  A. Rack\aff{2},
 \and O. Supponen\aff{1}
 }

\affiliation{\aff{1}Institute of Fluid Dynamics, Department of Mechanical and Process Engineering, ETH Z{\"u}rich, Sonneggstrasse 3, 8092 Z{\"u}rich, Switzerland
\aff{2}ESRF - The European Synchrotron, CS 40220, 38043 Grenoble Cedex 9, France}

\begin{document}
\maketitle

\begin{abstract}
Understanding how bubbles on a substrate respond to ultrasound is crucial for applications from industrial cleaning to biomedical treatments.
Under ultrasonic excitation, bubbles can undergo shape deformations due to Faraday instability, periodically producing high-speed jets that may cause damage.
While recent studies have begun to elucidate this behaviour for free bubbles, the dynamics of wall-attached bubbles is still largely unexplored.
In particular, the selection and evolution of non-spherical modes in these bounded systems have not previously been resolved in three dimensions, and the resulting jetting dynamics have yet to be compared with those observed in free bubbles.
In this study, we investigate individual micrometric air bubbles in contact with a rigid substrate and subjected to ultrasound. 
We introduce a novel dual-view imaging technique that combines top-view bright-field microscopy with side-view phase-contrast X-ray imaging, enabling visualisation of bubble shape evolution from two orthogonal perspectives.
This setup reveals the progression of bubble shape through four distinct dynamic regimes: purely spherical oscillations, onset of harmonic axisymmetric meniscus waves, emergence of half-harmonic axisymmetric Faraday waves, and the superposition of half-harmonic sectoral Faraday waves. 
This stepwise evolution contrasts with the behaviour of free bubbles, which  exhibit their ultimate Faraday wave pattern immediately upon instability onset.
For the substrate chosen, the resulting shape mode spectrum appears to be degenerate and exhibits a continuous range of shape mode degrees, in line with our theoretical predictions derived from kinematic arguments. 
While free bubbles also display a degenerate spectrum, their shape mode degrees remain discrete, constrained by the bubble spherical periodicity.
Experimentally measured ultrasound pressure thresholds for the onset of Faraday instability agree well with classical interface stability theory, modified to incorporate the effects of a rigid boundary.
Complementary 3D boundary element simulations of bubble shape evolution align closely with experimental observations, validating this method's predictive capability.
Finally, we determine the acceleration threshold at which shape mode lobes initiate cyclic jetting. Unlike free bubbles, jetting in wall-attached bubbles consistently emerges from the side not restricted by the substrate.

\end{abstract}


\section{Introduction}

When a liquid interface undergoes periodic oscillations with sufficient amplitude, standing wave patterns are generated.
This phenomenon is known as Faraday instability, named after its discoverer \cite{Faraday1831XVII.Surfaces}, who first observed its occurrence on the free surface of a vertically vibrating liquid bath.
Faraday also found that the resulting wave patterns oscillate at half the frequency of the driving force.
This type of response is denoted as half-harmonic, and is a hallmark of this kind of instability.
Faraday instability is also referred to as parametric instability because it involves the variation of a key parameter in the dynamical system—in this case, the effective gravitational acceleration.
Faraday's findings were later validated by \cite{Rayleigh1883VII.Support,Rayleigh1883XXXIII.Vibrations}.
The first linear stability analysis of Faraday waves for an inviscid liquid with a flat surface was performed by \cite{Benjamin1954TheMotion}, who demonstrated that the wave evolution follows a Mathieu equation.
This stability analysis shows that wave frequencies can be half-harmonic, harmonic, ultra-harmonic (integer multiples of half the driving frequency), or super-harmonic (integer multiples of the driving frequency).
However, the extension of this analysis to viscous fluids by \cite{Kumar1994ParametricFluids} revealed that the onset of the half-harmonic response requires the lowest acceleration, making it the only response that manifests.
Weak nonlinear effects can lead to interactions between different surface wave modes, influencing the selection of the dominant wave pattern.
\cite{Fauve1998PatternInstabilities} demonstrated that in an axisymmetric vessel, the most unstable linear eigenmode is an axisymmetric standing wave.
However, this axisymmetric pattern is nonlinearly unstable, and nonlinear effects ultimately favour the emergence of non-axisymmetric patterns.
These patterns can include lattices of stripes, squares, or hexagons \citep{Kudrolli1996LocalizedWaves}, as well as more complex structures such as quasi-patterns \citep{Christiansen1992OrderedWaves, Edwards1994PatternsExperiment}, triangles \citep{Muller1993PeriodicExperiment}, superlattice patterns \citep{Kudrolli1996LocalizedWaves, Wagner1999FaradayLiquid}, oscillons \citep{Arbell2000TemporallyFluids}, and rhomboidal patterns \citep{Arbell2000Two-modeWaves}.
The influence of surfactants on the interface of liquids with arbitrary depth was investigated by \cite{Kumar2002ParametricallyLiquids,Kumar2004OnLiquid} and extended to the specific case of thin liquid layers by \cite{Kumar2002InstabilityLayers,Matar2004NonlinearFilms}.

Faraday instability on spherical fluid interfaces was first examined by Lord \cite{Kelvin1863XXVIII.Liquid} and later by Lord \cite{Rayleigh1879VI.Jets}.
A key feature of the Faraday instability is the direct relationship between a specific pattern wavenumber and its frequency, as described by a dispersion relation.
In the case of a spherical interface, this relationship was derived by \cite{Rayleigh1879VI.Jets} and \cite{Lamb1932Hydrodynamics}, by considering the restoring force due to surface tension in a system of two incompressible, inviscid fluids, where one fluid entirely envelops the other, and reads:
\begin{equation}\label{eq:DispRel}
\omega_{0,l}^2 = \frac{l(l-1)(l+1)(l+2)}{(l+1)\rho_{\rm i} + l \rho_{\rm e}} \frac{\sigma}{R_0^3}, \quad \text{for } l = 2, 3, 4,...,
\end{equation}
where $\omega_{0,l}$ is the angular frequency of the pattern, $l$ is its angular wavenumber, $\rho_{\rm i}$ and $\rho_{\rm e}$ are the densities of the internal and external fluids, respectively, $\sigma$ represents surface tension, and $R_0$ is the radius of the spherical interface between the two fluids.
Due to the spherical symmetry, the wavenumber must take on integer values. 
Consequently, when the driving frequency $\omega_{\rm d} = 2\omega_{0,l}$ corresponds to a non-integer wavenumber according to the dispersion relation (\ref{eq:DispRel}), the actual wavenumber adjusts to the nearest integer, leading to a \emph{quantised} response spectrum.
However, in such cases, the onset acceleration required to generate a pattern will exceed the value necessary when the driving frequency directly resonates with an integer wavenumber.

\begin{figure} 
    \centering
        \includegraphics[width=\columnwidth]{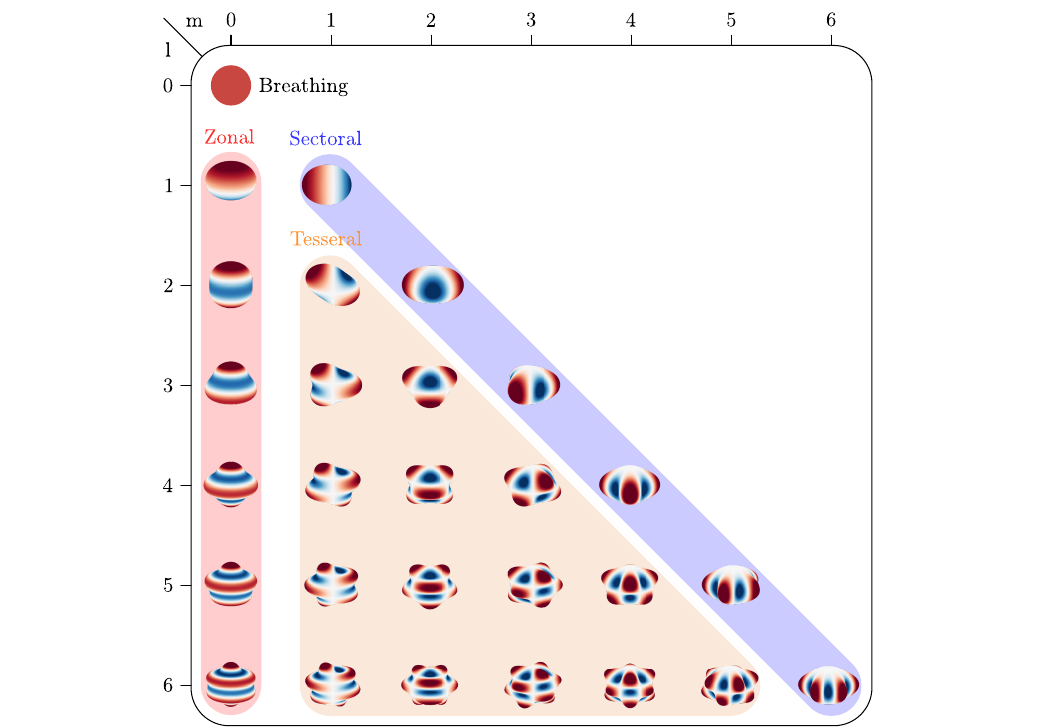}
    \caption{
    Shape modes of a spherical interface ordered by degree $l$ and order $m$, up to the sixth degree.
    Zonal modes occupy the column marked in red ($m=0$), sectoral modes are located along the highlighted diagonal in blue ($l=m$), and tesseral modes are distributed across the remaining yellow area ($l\neq m$).
    The $l=0$ mode corresponds to a uniform spherical deformation and is commonly referred to as breathing mode.}  
    \label{fig:ShapeModesIllustration}
\end{figure}

The interface deformation, $\eta(\theta, \phi, t)$, is classically expressed using real spherical harmonics, $Y_{l}^{m}(\theta, \phi)$, as follows:
\begin{gather}\label{eq:DefSphHarm}
\eta(\theta,\phi,t) = \sum_{m=0}^{l} a_l^{m}(t) Y_{l}^{m}(\theta,\phi), \\
Y_{l}^{m}(\theta,\phi) = N_{l}^{m} P_{l}^{m} (\cos (\theta)) \cos(m \phi), \quad -\pi \le \theta \le \pi, \quad 0 \le \phi \le 2\pi, 
\end{gather}
where $\theta$ and $\phi$ are the polar and azimuthal angles in spherical coordinates, respectively.
$P_{l}^{m}(\cos (\theta))$ denotes the associated Legendre function of degree $l$ and order $m$.
$N_{l}^{m} = (\max(P_{l}^{m} (\cos (\theta)) \cos(m \phi)))^{-1}$ is a normalisation factor that ensures the value of the spherical harmonic is constrained to unity.
$a_l^{m}(t)$ represents the time-dependent amplitude of each spherical harmonic.
By normalising the harmonics in this way, the shape mode magnitudes, $a_l^m(t) Y_l^m$, can be directly compared through the coefficients $a_l^m(t)$ alone.
Spherical harmonics are also known as \emph{shape modes} and are ordered by their wavenumber pairs [$l, m$].
For a given $l \neq 0$, these modes are classified as follows: \emph{zonal} modes when $m = 0$, \emph{sectoral} modes when $m = l$, and \emph{tesseral} modes when $m \neq 0$ and $m \neq l$.
Figure \ref{fig:ShapeModesIllustration}, which illustrates modes up to the sixth degree, shows that zonal modes are axisymmetric with $l$ azimuthal nodal lines, sectoral modes exhibit star symmetry with $m$ polar nodal lines, and tesseral modes have ($l - m$) azimuthal nodal lines along with $m$ polar nodal lines.
The $l=0$ mode represents a uniform spherical deformation, often referred to as \emph{breathing mode}.
It characterises radial oscillations of bubbles driven by a sinusoidal acoustic field.
The $l=1$ mode manifests as an alternating translational motion and must correspond to a gravity wave rather than a capillary wave, as it does not deform the interface and thus cannot be restored by surface tension.
It is crucial to note that the dispersion relation (\ref{eq:DispRel}), which describes the linear stability of the interface, is independent of the spherical harmonic order $m$.
Consequently, a departure from the spherical symmetry state leads to a set of $l+1$ linearly independent and equally probable solutions.
In this context, the spectrum is termed as \emph{degenerate}.
The final deformation pattern, determined by a specific combination of shape modes with the same $l$ as represented in equation (\ref{eq:DefSphHarm}), will be selected by nonlinear effects \citep{Chossat1991Steady-State03-Symmetry}.

Drops and bubbles are two of the most common forms of fluids encapsulated within one another.
The predictions of the Rayleigh-Lamb spectrum for small shape oscillations of liquid drops have been experimentally confirmed by \cite{Trinh1982AnSystems} by suspending the drops in a host fluid through acoustic forces.
To eliminate the influence of acoustic trapping and enable fully non-invasive investigations, researchers have also conducted experiments in space using free-floating drops \citep{Wang1996OscillationsSpace} or during parabolic flights employing a liquid layer enclosed within a spherical container \citep{Falcon2009CapillaryGravity}.
Nevertheless, the applicability of the Rayleigh-Lamb spectrum is limited to inviscid fluids and infinitesimal interfacial deformations.
The influence of viscosity has been studied by  \cite{Miller1968TheFluid}, \cite{Prosperetti1980FreeProblem} and \cite{EboAdou2016FaradayAnalysis}, who found that increased viscosity raises the acceleration threshold for the onset of shape modes, particularly as the wave number increases.
Concerning finite deformations,  \cite{Tsamopoulos1983NonlinearBubbles} were the first to analytically investigate weakly nonlinear effects on inviscid axisymmetric drops.
Later, \cite{Lundgrent1988OscillationsEffects} employed a boundary integral method to compute large axisymmetric motions with weak viscous effects.
More recently, \cite{Ebo-Adou2019FaradaySimulation} conducted fully three-dimensional viscous simulations, which revealed the final deformation patterns of the interface as determined by nonlinear effects.
The $l = 2$ mode alternates between oblate and prolate spheroidal shapes, and modes with $l \geq 3$ produce patterns analogous to Platonic solids.
Building on this, \cite{Panda2024DropDrops} extended the study to higher spherical harmonic degrees and amplitudes.

The Rayleigh-Lamb spectrum, originally developed for free spherical interfaces, has been extended to also account for constrained spherical inviscid drops. \cite{Strani1984FreeSupport} first introduced this extension by considering a constraint in the form of a spherical bowl.
Later, \cite{Bostwick2014DynamicsTheory} explored the dynamics of an inviscid sessile drop resting on a flat, planar support, analysing its natural oscillations.
They showed that these oscillations are influenced by the static contact angle and the mobility of the contact line.
Notably, they also found that a flat support breaks the spectral degeneracy, leading to shape modes with the same degree $l$ but different order $m$ exhibiting distinct resonance frequencies.
These theoretical predictions align reasonably well with experimental observations by \cite{Chang2013SubstrateDrops,Chang2015DynamicsExperiment} for pinned drops with minimal gravitational effects, although the agreement weakens for flatter drops and higher-order modes.
The researchers also discovered the phenomenon of mode mixing, where multiple mode shapes are excited by a single driving frequency.
An alternative analytical approach for predicting the shape mode spectrum of sessile drops, proposed by \cite{Sharma2021OnProblem}, provides more accurate results for flatter drops.
In their experimental study, \cite{Vukasinovic2007DynamicsVibration} found that harmonic axisymmetric standing waves emerge even under minimal forcing.
When the forcing amplitude exceeds a critical threshold, the Faraday instability generates half-harmonic azimuthal waves along the contact line.
At even higher forcing amplitudes, these waves merge with the harmonic waves, creating a lattice-like pattern.
These distinctive patterns were also reported in three-dimensional viscous simulations by \cite{Panda2023AxisymmetricDrop}.

Compared to drops, bubbles exhibit richer dynamics owing to the additional degree of freedom introduced by the compressibility of the inner fluid.
When subjected to an acoustic driving with a wavelength significantly larger than the bubble size, free bubbles undergo spherical oscillations, which correspond to the $l=0$ mode in spherical harmonics terminology.
The natural angular frequency $\omega_{0,0}$ of this mode, for small deformations, can be determined by linearising the Rayleigh-Plesset equation in combination with the polytropic gas approximation, which describes the dynamics of a spherical bubble \citep{Plesset1977BubbleCavitation}, as follows:
\begin{equation}
\omega_{0,0}^2 = \frac{1}{\rho_l R_0^2}\left(3n p_{\infty} + \frac{2 (3n-1) \sigma}{R_0} \right),
\end{equation}
where $\rho_l$ is the density of the liquid medium, $n$ is the gas polytropic index, and $p_{\infty}$ is the ambient pressure.
In addition to shape oscillations induced by direct forcing, as seen with drops, bubbles can experience shape oscillations induced by the harmonic radial acceleration stemming from their spherical oscillations.
To predict the amplitude of shape oscillations in the context of inviscid fluids, a classical approach involves employing a second-order ordinary differential equation linearised with respect to the mode amplitude \citep{Plesset1954OnSymmetry,Benjamin1958ExcitationWork,Hsieh1961OnBubbles, Eller1970InstabilityField}.
The corresponding problem for viscous fluids has been treated by \cite{Prosperetti1977ViscousFlows,Ceschia1978OnLiquid,Brenner1995BubbleSonoluminescence,Hilgenfeldt1996PhaseBubbles}.
By performing a first-order analysis, \cite{Francescutto1978PulsationBubbles} derived a threshold condition for the onset of shape modes on an oscillating bubble in a sound field.
\cite{Versluis2010MicrobubbleDriving} leveraged these analytical findings to compare the theoretical acoustic pressure thresholds for the onset of shape modes against experimental results obtained from testing free micrometric bubbles driven by ultrasound, finding a good agreement.

However, these studies do not account for the nonlinear impact of a specific shape mode on the breathing mode, bubble translation, and other shape modes.
Ignoring this influence in the aforementioned models results in predictions of unbounded growth for any shape deformation.
As a result, the applicability of these models is limited to predicting the amplitude threshold for the onset of shape oscillations.
Initially, the coupling between the breathing mode and a single shape mode has been studied analytically by \cite{Longuet-Higgins1989MonopoleModes,Longuet-Higgins1989MonopoleProblem, Longuet-Higgins1991ResonanceOscillations,Mei1991ParametricBubble, Feng1992OnOscillations,Feng1994BifurcationResonance,Feng1997NONLINEARDYNAMICS} and shown to be both periodic and energy conserving. 
Building on this, subsequent research has progressively advanced the understanding of the interaction between the breathing mode, translational motion, and multiple shape modes of a bubble \citep{Feng1995TranslationalOscillations,Doinikov2004TranslationalOscillations,Shaw2006TranslationDeformation,Shaw2009TheWave,Shaw2017NonsphericalCoupling}.
Experimental studies with acoustically trapped micrometric bubbles driven by ultrasound carried out by \cite{Guedra2016ExperimentalMicrobubbles, Guedra2017DynamicsThreshold} have confirmed these theories, reporting strong nonlinear mode coupling, including the saturation of instability and the triggering of nonparametric shape modes.
The development of nonlinear coupled models has been pivotal in elucidating the mechanisms behind the sound emission stemming from the activation of the breathing mode \citep{Minnaert1933XVI.Water} and the intriguing ``dancing motion" of a bubble in an acoustic field \citep{Kornfeld1944OnCavitation}.
\cite{Cattaneo2025CyclicDelivery} experimentally characterised the pattern for the first six shape modes of ultrasound-driven phospholipid-coated microbubbles in contact with a supersoft, superhydrophilic substrate.
This configuration minimises the effects of wall confinement, allowing the bubbles to behave as if they are in an unbounded fluid.
The shape mode patterns found correspond to the numerical predictions from \cite{Ebo-Adou2019FaradaySimulation} for droplets, suggesting a common interfacial response to both drops and bubbles, at least when the two cases involve similar density and viscosity ratios, as well as comparable surface tension.

While we have accumulated substantial knowledge regarding the oscillations of bubbles in unbounded fluids, our understanding of how confining surfaces influence bubble dynamics remains relatively limited. 
\cite{Strasberg1953TheLiquids} was the first to quantify the influence of an infinitely rigid wall on the breathing mode resonance of a single bubble using the method of images. \cite{Shklyaev2008LinearSubstrate,Fayzrakhmanova2011BubbleHysteresis} explored the theoretical aspects of linear natural and forced shape oscillations of a hemispherical bubble on a solid substrate, examining the effects of bubble compressibility and contact angle hysteresis.
Their findings revealed that the contact line dynamics induces an interaction between the breathing mode and the shape modes, even within a linear framework.
\cite{Prosperetti2012LinearSurfaces, Vejrazka2013LinearDrop} investigated the shape modes of a spherical bubble pinned to a ring, while \cite{Ding2022OscillationsBubble} analysed the hydrodynamic stability of a bubble on a solid substrate.
This study uncovered a complex relationship between the frequency spectrum and wetting properties, which is influenced by factors such as the static contact angle, the equilibrium bubble pressure, and the contact-line dynamics, and showed that the constraint removes the mode degeneracy seen in free bubbles.
Finally, \cite{Fauconnier2020NonsphericalBubble} investigated experimentally the modal behaviour of a wall-attached bubble under acoustic excitation.
Through the analysis of top-view recordings using spherical harmonics decomposition, the study suggested the absence of mode degeneracy, as well as the presence of modal mixing and competition.

\begin{figure} 
    \centering
        \includegraphics[width=\columnwidth]{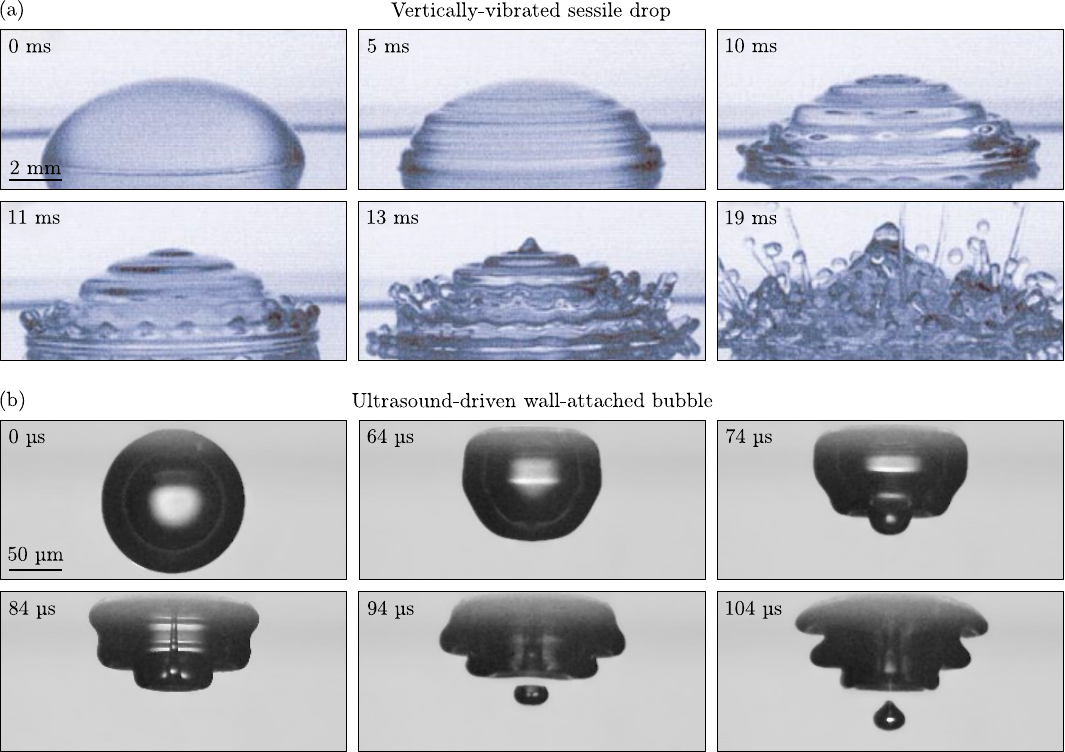}
    \caption{(a) Sessile water droplet with a \SI{5}{\milli\meter} radius subjected to vertical vibrations at a frequency of \SI{1080}{\hertz}. 
    Reproduced with permission from the work by \cite{Vukasinovic2000Vibration-InducedAtomization}.
    (b) Wall-attached air bubble with a \SI{68}{\micro\meter} radius subjected to ultrasound driving at a frequency of \SI{100}{\kilo\hertz}.
    Reproduced with permission from the work by \cite{Cattaneo2023JetSubstrate}.
    }
    \label{fig:Introduction_ShapeModeSequences}
\end{figure}

When the interface acceleration surpasses a certain threshold, the amplitude of Faraday waves rapidly increases, causing the wave profile to become unstable.
This instability ultimately leads to the formation of a jet that moves from the higher-density fluid to the lower-density fluid.
\cite{Longuet-Higgins1983BubblesSurface} was the first to observe this phenomenon on a vertically vibrating flat water-air surface.
In some remarkable cases, he observed jets produced by a relatively mild excitation amplitude of \SI{0.5}{\milli\meter} that rose up to a height of \SI{1.7}{\meter}.
The understanding of Faraday waves-induced jets on flat interfaces was significantly advanced by the work of \cite{Goodridge1996ThresholdWaves, Goodridge1997ViscousWaves, TaoShi1997BreakingState, Hogrefe1998Power-lawWaves, Goodridge1999BreakingRates, Zeff2000SingularitySurface}. 
These studies identified the critical wave height necessary for jet formation and demonstrated that the surface undergoes a self-similar collapse prior to jetting.
\cite{Vukasinovic2000Vibration-InducedAtomization,Vukasinovic2001ModeVibration,James2003Vibration-inducedBursting} were the first to observe Faraday waves-induced jetting in a spherical sessile water drop placed on a vibrating plate.
A sequence of images from their studies on drop jetting is reproduced in figure \ref{fig:Introduction_ShapeModeSequences}(a).
The jets, which extended outward from the droplet, led to atomisation and drop bursting through Rayleigh-Plateau instability.
Recently, \cite{Prasanna2024SynchrotronAtomization} studied viscoelastic drops and demonstrated that the high resistive stresses in the jets significantly enhance their stability against the Rayleigh–Plateau instability.
\cite{Cattaneo2023JetSubstrate} provided visualisations that clearly depict the formation of cyclic jets fostered by the Faraday instability on wall-attached bubbles oscillating under ultrasound driving.
Figure \ref{fig:Introduction_ShapeModeSequences}(b) presents a series of images capturing this process.
We believe that cyclic jets driven by Faraday instabilities, though previously unrecognised, have likely been observed in multiple earlier studies employing driving frequencies spanning from hertz \citep{Crum1979SurfaceBubbles}, through kilohertz \citep{Prabowo2011SurfaceBubbles}, and up to megahertz \citep{Vos2011}.
Shape-mode induced jets were also observed for bubbles in contact with a vibrating surface, as reported by \cite{Biasiori-Poulanges2023SynchrotronCavitation}.
Recently, \cite{Dhote2024StandingBubble} successfully used Gerris, an open-source computational fluid dynamics software, to numerically simulate and demonstrate the formation of a shape mode-induced jet in an inviscid, incompressible, pre-deformed bubble.
Finally, in our previous work \citep{Cattaneo2025CyclicDelivery}, we studied the jet formation driven by Faraday instability on practically unconstrained, phospholipid-coated, microbubble ultrasound contrast agents subjected to high-frequency ultrasound, revealing a multi-jetting behaviour, where the number of jets scales with the number of lobes in the shape mode.
We also discovered that these jets are capable of piercing the cell membrane, enabling precise and targeted drug delivery.

While significant progress has been made in understanding oscillations and jetting in acoustically-driven free bubbles, the dynamics of wall-attached bubbles remain poorly understood.
In particular, the selection and evolution of shape mode patterns in these bubbles have not been experimentally resolved in three dimensions, and the resulting shape mode-induced jetting has yet to be compared with that observed in free bubbles.
The present study addresses this gap by examining bubbles, with radii around \SI{100}{\micro\meter}, in contact with a rigid substrate, subjected to low-frequency ultrasound.
Building on the work of \cite{Fauconnier2020NonsphericalBubble}, which was limited to top-view imaging, we introduce a dual-imaging technique that combines bright-field microscopy with phase-contrast X-ray synchrotron imaging.
This approach captures bubble dynamics from two orthogonal perspectives, providing visual access to the full 3D bubble shape.
The X-ray modality minimises refraction artefacts and reveals all bubble folds, including those on the distal side that are hidden in traditional back illumination.
This dual-view methodology enables, for the first time, direct observation of the formation, spectrum, and temporal evolution of Faraday instability leading up to jet formation in wall-attached bubbles.
The results allow quantitative comparison with theoretical predictions and 3D numerical simulations, offering new insights into the mechanisms driving these complex phenomena.
This paper is structured as follows.
Section \ref{Sec2} outlines our experimental setup that combines X-ray synchrotron and conventional light sources to enable dual-view imaging.
Section \ref{Sec3} outlines the theoretical framework for describing bubble deformations in contact with a wall and explores the permissible shape modes as a function of the wetting conditions.
Section \ref{Sec4} provides an overview of the bubble response, breaking it down into four sequential regimes.
Section \ref{Sec5} presents a quantitative experimental analysis of the four regimes, supplemented by comparisons with theoretical predictions and numerical simulations, and offers physical interpretations of the findings.
Section \ref{Sec6} provides a comparison between the experimental visualisations and three-dimensional simulations of bubble dynamics generated using the boundary element method.

\section{Experimental setup} \label{Sec2}
\begin{figure} 
    \centering
        \includegraphics[width=0.95\columnwidth]{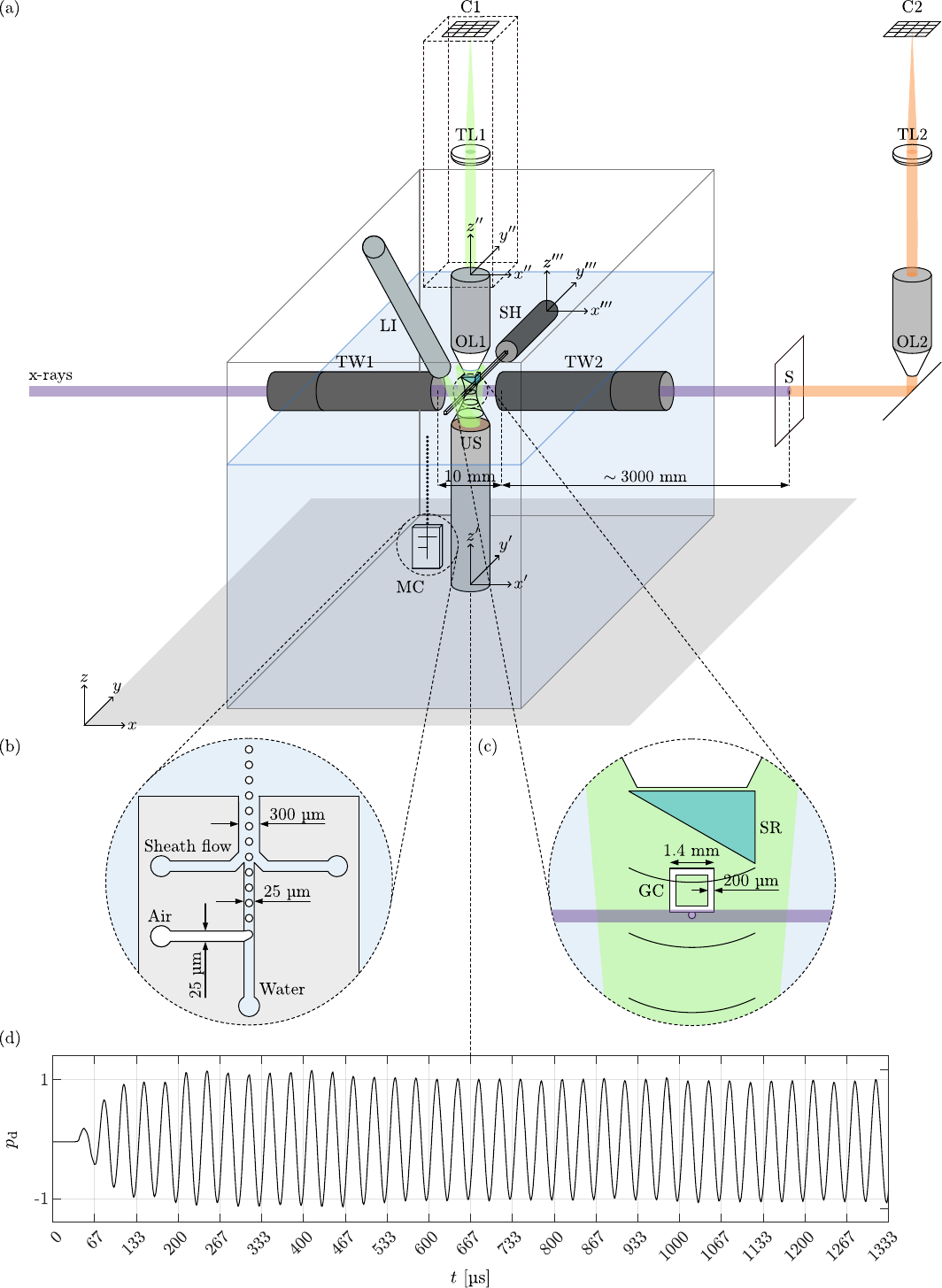}
    \caption{(a) Schematics of the experimental setup. (C1-C2) Cameras, (GC) Glass capillary, (LI) LED illuminator, (MC) Microfluidic chip, (OL1-OL2) Objective lenses, (S) Scintillator, (SH) Sample holder, (SR) Sound reflector, (TL1-TL2) Tube lenses, (TW1-TW2) Telescopic windows, (US) Ultrasound transducer.
    (b) Geometry of the microfluidic bubble-generator chip.
    A single bubble is diverted from the upward bubble stream and propelled towards the bottom of the glass capillary using a manually operated syringe.
    (c) Testing conditions with a single microbubble positioned underneath a glass capillary.
    (d) Acoustic driving pressure produced by the ultrasound transducer and measured by a needle hydrophone, normalised to the steady-state amplitude value.}
    \label{fig:Setup}
\end{figure}

A schematic of the experimental setup is shown in figure \ref{fig:Setup}(a).
In a water bath $(T_l\approx \SI{22}{\celsius} )$ filled with deionised water, a rising stream of monodisperse bubbles with an equilibrium radius within the range $R_0=\SIrange{60}{140}{\micro\meter}$ is generated using a PDMS (polydimethylsiloxane) microfluidic chip fabricated using soft lithography.
The chip layout consists of a T-junction where the channel conveying the continuous phase (water) intersects with the channel transporting the dispersed phase (air), resulting in the formation of bubbles.
Additionally, two channels further downstream create a sheath flow which assist the bubble separation by increasing the flow rate. 
The geometry of the microfluidic chip is depicted in figure \ref{fig:Setup}(b). 
The liquid phase is supplied by a syringe pump (NE-300, Darwin Microfluidics) with a flow rate of $\SI{50}{\micro\liter\per\minute}$. 
The gas is fed by an air compressor (Fatmax DST 101/8/6, Stanley) and its pressure, which determines the bubble size, is controlled with a pressure reducing valve (RP1000-8G-02, CKD). 
Finally, the sheath flow is provided by a syringe pump (NE-300, Darwin Microfluidics) with a combined flow rate of \SI{200}{\micro\liter\per\minute}.

One of these micrometric air bubbles is diverted from the stream using a manually operated syringe and placed to rest on the bottom of a square borosilicate glass capillary immersed in water. 
A square capillary, characterised by a side length $l$ of \SI{1.4}{\milli\meter} and a wall thickness $\delta$ measuring \SI{0.2}{\milli\meter} (8100-50, CM Scientific), is chosen over a solid substrate because it mitigates sound reflections due to its hollow structure filled with water and capillary walls significantly thinner than the wavelength $\lambda$ of the acoustic driving ($ \lambda / \delta = 250$).
Nevertheless, its side length is sufficient large to avoid boundary effects ($l/ 2R_0 > 5$).
Moreover, the capillary cross section provides good bending rigidity to the substrate, preventing unintended oscillations when subjected to acoustic stimulation.
Figure \ref{fig:Setup}(c) presents a close-up view of the bubble in contact with the capillary.
The bubble is acoustically driven at a frequency $f_{\rm d} = \SI{30}{\kilo\hertz}$ with pressure amplitudes in the range $p_{\rm a}=\SIrange{1}{15}{\kilo\pascal}$ using an ultrasound transducer (GS30-D25, The Ultran Group) positioned in the water bath perpendicular to the horizontal plane.
The driving pulse is generated using a function generator (LW420B, Teledyne LeCroy) and subsequently amplified by a radiofrequency power amplifier (1020L, E\&I).
The resulting ultrasound pulse shape, as measured by a needle hydrophone ($\SI{0.2} {\milli\meter}$, NH0200, Precision Acoustics), is depicted in figure \ref{fig:Setup}(d).

Side-view visualisations of the bubble response are acquired using high-speed synchrotron X-ray phase-contrast radiography at the ID19 beamline of the European Synchrotron Radiation Facility (ESRF), during the $7/8 +1$ filling mode with an integral storage ring current of $\SI{200}{\milli\ampere}$.
We employ X-ray synchrotron radiation because it is far less susceptible to refraction on curved surfaces compared to visible light.
This choice helps eliminate the black bands typically seen on spherical surfaces illuminated with visible light (as observed in Figure \ref{fig:Introduction_ShapeModeSequences}(b)).
These bands can obscure the surface evolution of the bubble during lobe folding and hinder the early detection of jet formation.
Moreover, the quasi-parallel X-ray beam at the ID19 beamline, enabled by its exceptionally long source-to-sample distance, effectively minimises source size effects (penumbral blur), enabling us to clearly resolve all interface folds, even on the distal side of the bubble.
This enhanced clarity makes it easier to identify the shape modes that develop, particularly when the bubble loses its axisymmetry.
X-ray light requires the use of indirect detectors that first convert X-rays into visible light and then into electrons.
However, these detectors lose efficiency as they approach the optical resolution limit, necessitating ultra-thin scintillators for effective performance.
As a result, high-speed X-ray imaging often comprises spatial resolution to enhance the signal-to-noise ratio, leading to less sharply defined interfaces compared to those commonly seen in visible light imaging.

For this study, the bubble in contact with the glass capillary is illuminated by a pink X-ray beam generated by the U17.6 undulator of the beamline, which is set with a \SI{12}{\milli\meter} gap (undulator period $\lambda_{\rm x} = \SI{17.6}{\milli\meter}$, number of periods of magnet pairs $N_{\rm x} = 92$).
The insertion device produces an on-axis central beam energy (first harmonic) at \SI{17.9}{\kilo\electronvolt}, with a $2\%$ bandwidth (FWHM), while the second and third harmonics are suppressed by two orders of magnitude in photon flux.
The beam is filtered with a set of mandatory optical elements along the vacuum flight path: a \SI{1.8}{\milli\meter}-thick diamond window, a \SI{0.7}{\milli\meter}-thick aluminum filter, and successive thin carbon and beryllium windows.
The beam is collimated to the field of view by using eight compound refractive lenses (CRLs) from the beamline transfocator, located approximately \SI{35}{\meter} from the source, acting as beam condensers. 
Additionally, two sets of in-vacuum slits are employed to crop the beam to match the field of view. 
The indirect detector consists of a \SI{250}{\micro\meter}-thick LuAG:Ce scintillator (Crytur), coupled via a mirror to a microscope objective lens of focal length  $\mathcal{F} = \SI{10} {\milli\meter}$ (MY20X-804, Mitutoyo) with lead glass protection and a $\mathcal{F} = \SI{200} {\milli\meter}$ tube lens for a total magnification of $20\times$.
The luminescent image is then relayed to a high-speed camera (HPV-X2, Shimadzu).
The distance between the sample and the scintillator is around \SI{3}{\meter}.

The water tank is constructed from PMMA (polymethylmethacrylate) and features two water-tight telescopic windows on two opposite facing walls comprising aluminium tubes ($\varnothing$0.5'' stackable tube lens, Thorlabs) and circular \SI{0.5}{\milli\meter}-thick PMMA windows secured at their ends by retaining rings and sealed with high-vacuum grease. 
These telescopic windows are designed to minimise the path traversed by X-ray radiation in water, aiming to decrease absorption and consequently enhance the signal.
Therefore, the distance between the windows is minimised to the greatest extent, compatible with spatial limitations dictated by other components, at approximately $d= \SI{10}{\milli\meter}$. 
The transmission coefficient of the X-ray signal, pertaining to the thickness of the water layer utilised, can be estimated to be approximately equal to $T=0.4$ using the relation $T = e^{- \epsilon_{\rm l} \rho_{\rm l} d}$  \citep{Henke1993X-ray1-92}, where $\epsilon_{\rm l}$ is the mass absorption coefficient, which for water at $\SI{17.9}{\kilo\eV}$ is measured to be $\SI{9.263e-2}{\metre\squared\per\kilo\gram}$ \citep{Hubbell1995TablesInterest} and $\rho_{\rm l}$ is the water density.

Simultaneous top-view visualisations of the bubble response are captured using high-speed visible-light microscopy.
A custom-built upright microscope is realised for the purpose using modular optomechanics components (Thorlabs, cage system) and equipped with a water-dipping  objective lens of focal length  $\mathcal{F} = \SI{20} {\milli\meter}$ (N10XW-PF, Nikon) and a $\mathcal{F} = \SI{400} {\milli\meter}$ tube lens (TL400-A, Thorlabs) for a total magnification of $20\times$. 
Backlight illumination is provided by a custom-built high-power LED illuminator used in continuous mode for live imaging and in burst mode for video recording. 
The light emission is guided with a liquid optical fibre and sent through the sample and then into the objective by reflecting it against the upper surface of the ultrasound transducer.
Sound reflections against the objective lens are mitigated by attaching a transparent PMMA prism to its base (see figure~\ref{fig:Setup}(c)).
Video recordings are captured using a high-speed camera (HPV-X2, Shimadzu).

The top-view and the side-view camera are synchronised ($<\SI{10}{\nano\second}$ of delay between the two) and allow for recordings at a frame rate of \SI{150}{\kilo\hertz} over 256 frames of continuous visualisation of a \numproduct{640 x 400} \unit{\micro\meter} field of view with a 1.6-\unit{\micro\meter} pixel resolution.  
The activation of the ultrasound pulse, cameras recording, light flash, and X-ray beam switching are synchronised through a delay generator (DG645, Stanford Research Systems).

Due to X-rays having a shorter wavelength than visible light, radiographs exhibit a remarkably high depth of field.
While this ensures sharp focus across all image planes, it impedes gaugeing the distance of an object from the optical window through distance blur.
Consequently, positioning the bubble under study within the acoustic focal volume requires the combined use of side and top views.
The one-time alignment protocol implemented to ensure the coincidence of the centres of the fields of view and the acoustic focal point is as follows:
first, we align the X-ray optical path within the telescopic windows by repositioning the optical table supporting the setup.
Next, we position the needle hydrophone parallel to the horizontal plane passing through the X-ray optical path, halfway between the two telescopic windows, and align its tip with the centre of the X-ray field of view through a motorised three-axis microtranslation stage (PT3/M-Z8, Thorlabs).
We manoeuvre the ultrasound transducer using a manual three-axis microtranslation stage (three DTS50/M, Thorlabs) to obtain the maximum signal from the hydrophone. 
We position the entire top-view videomicroscopy system by means of a manual three-axis microtranslation stage (\#12694, \#66-511, Edmund Optics) so that the tip of the hydrophone is in the centre of the microscope field of view and in focus.
Finally, the hydrophone is replaced by the square capillary.

\section{Theoretical framework}\label{Sec3}
\subsection{Kinematic model}

\begin{figure} 
    \centering   \includegraphics[width=\columnwidth]{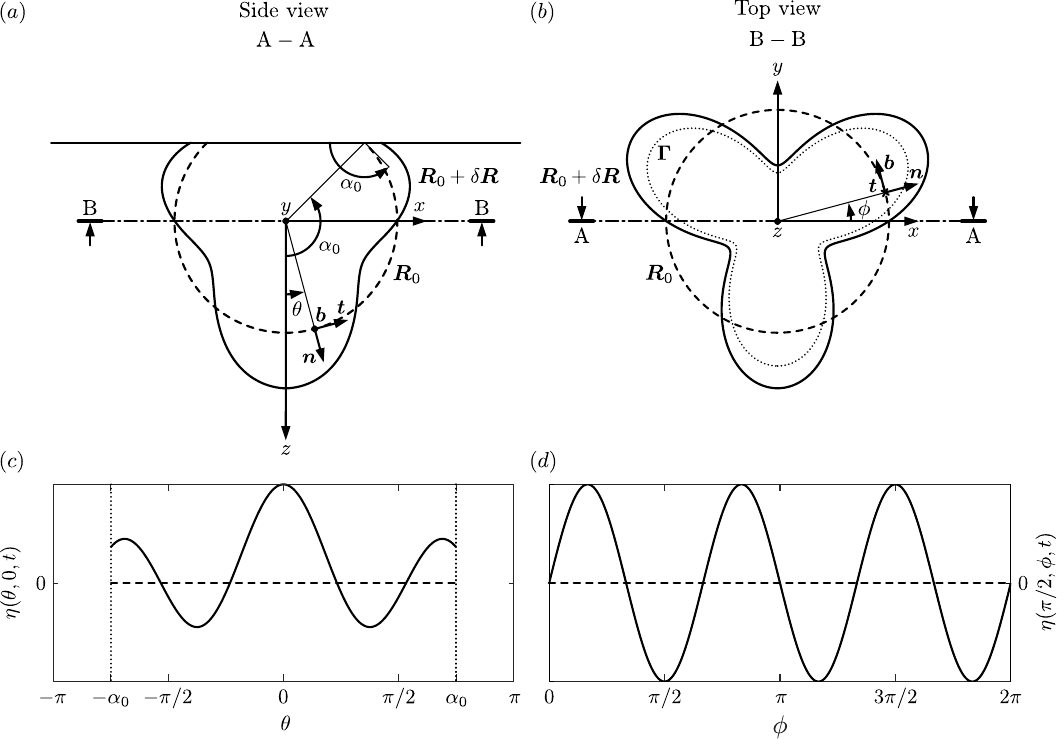}
    \caption{Definition sketch of a bubble resting against a rigid flat substrate with a static contact angle $\alpha_0$.
    (a) and (b) depict the cross-sections along the planes defined by projection lines $\text{A}-\text{A}$ and $\text{B}-\text{B}$, respectively.
    The equilibrium surface, represented by a dashed line, is defined as ${\boldsymbol{R}}_0(\theta,\phi)$, where polar angle $\theta \in [-\alpha_0,\alpha_0]$ and the azimuthal angle $\phi \in [0, 2\pi]$ serve as surface coordinates.
    The time-dependent surface deformation is denoted as $\delta{\boldsymbol{R}}(\theta,\phi,t)$.
    The deformed surface ${\boldsymbol{R}}_0(\theta,\phi) + \delta{\boldsymbol{R}}(\theta,\phi,t)$ is depicted by a solid line.
    $\boldsymbol{n}$, $\boldsymbol{t}$ and $\boldsymbol{b}$ denote the normal, tangential and binormal unit vectors to the equilibrium surface, respectively.
    The contact line $\boldsymbol{\Gamma}$ is shown as a dotted line.
    (c) and (d) represent the surface deformation normal to the equilibrium surface $\delta{\boldsymbol{R}}(\theta,\phi,t) \cdot \boldsymbol{n} = \eta$ in the $\text{A}-\text{A}$ and $\text{B}-\text{B}$ plane, respectively.}
    \label{fig:ProblemDefinition}  
\end{figure}
Let us consider a bubble resting against a rigid flat substrate, as depicted in the side and top views in figure \ref{fig:ProblemDefinition}(a)-(b).
Given the small Bond number, $\text{Bo} = {(\rho_{\rm l} - \rho_{\rm g}) g R_0^2}/{\sigma} \sim O(10^{-3})$, where $(\rho_{\rm l} - \rho_{\rm g})$ is the density difference between water and air, and $g$ is the gravitational acceleration, the influence of gravitational effects is negligible for the problem at hand.
Consequently, under equilibrium conditions, the bubble can be approximated as a spherical cap with a static contact angle $\alpha_0$.
We define the origin of the reference frame at the geometric centre of the uncut sphere.
The surface of the bubble at equilibrium can be parametrically described as ${\boldsymbol{R}}_0(\theta,\phi)$, where $\theta \in [-\alpha_0,\alpha_0]$ is the polar angle and $\phi \in [0, 2\pi]$ is the azimuthal angle. 
By normalising the bubble equilibrium surface ($\|{\boldsymbol{R}}_0\| = 1$), the arc angle and arc length values are equal.
The surface deformation is defined as $\delta{\boldsymbol{R}}(\theta,\phi,t)$.
This deformation may displace the contact line, denoted as $\boldsymbol{\Gamma}(\theta)$, along the substrate.
The unit vector normal to  ${\boldsymbol{R}}_0$ is denoted by ${\boldsymbol{n}}$.
The unit vector tangential to ${\boldsymbol{R}}_0$ and normal to $\boldsymbol{\Gamma}$ is denoted by ${\boldsymbol{t}}$.
The unit vector tangential to ${\boldsymbol{R}}_0$ and to $\boldsymbol{\Gamma}$ is denoted by ${\boldsymbol{b}} = {\boldsymbol{n}}  \times {\boldsymbol{t}}$.
The unit vector normal to the substrate is denoted as ${\boldsymbol{n}}_{\rm s}$.
The bubble deformation can be decomposed into the component normal to the bubble equilibrium surface, defined as:
\begin{equation}
\delta_{\perp} {\boldsymbol{R}} = \left(\delta{\boldsymbol{R}} \cdot {\boldsymbol{n}} \right){\boldsymbol{n}} = \eta {\boldsymbol{n}},
\end{equation}
and the component tangential to the bubble equilibrium surface, represented by:
\begin{equation}
\delta_{\parallel} {\boldsymbol{R}} = \left(\delta{\boldsymbol{R}} \cdot {\boldsymbol{t}} \right){\boldsymbol{t}} = \tau {\boldsymbol{t}},
\end{equation}
as $\delta{\boldsymbol{R}} \perp {\boldsymbol{b}}$.
An example of normal deformation $\eta$ in the plane perpendicular to the substrate defined by the projection line $\text{A}-\text{A}$, and thus identified as $\eta(\theta,0,t)$, is shown in figure \ref{fig:ProblemDefinition}(c).
Instead, an example of normal deformation $\eta$ in the plane parallel to the substrate defined by the projection line $\text{B}-\text{B}$, and thus identified as $\eta(\pi/2,\phi,t)$, is shown in figure \ref{fig:ProblemDefinition}(d).

The equilibrium contact angle $\alpha_0$ is related to the surface tensions of the solid/liquid $\sigma_{\rm sl}$, solid/gas $\sigma_{\rm sg}$ and liquid/gas $\sigma_{\rm lg}$ through the Young-Dupré equation:
\begin{equation}\label{eq:YD}
({\boldsymbol{n}} \cdot {\boldsymbol{n}}_{\rm s})|_{\theta = \alpha_0} \equiv \cos(\alpha_0) = \frac{\sigma_{\rm sg} - \sigma_{{\rm sl}}}{\sigma_{\rm lg}}.
\end{equation}
A perturbation in the bubble surface $\delta{\boldsymbol{R}}$ results in a change in the wetting condition \citep{Myshkis1987Low-GravityMechanics, Tyuptsov1966HydrstaticsFields}:
\begin{equation}\label{eq:vwr}
 \partial_{\theta}\eta|_{\theta=\alpha_0} - \cot(\alpha_0)\eta|_{\theta=\alpha_0} = -\delta\alpha.
\end{equation}
A full derivation of this relation is provided in Appendix \ref{appA}.
We use this kinematic boundary condition to determine the spectrum of shape modes for the deformation $\delta{\boldsymbol{R}}$ under various wetting conditions.

\subsection{Shape modes for wall-attached bubbles}\label{Sec32}

\begin{figure}
    \centering
        \includegraphics[width=\columnwidth]{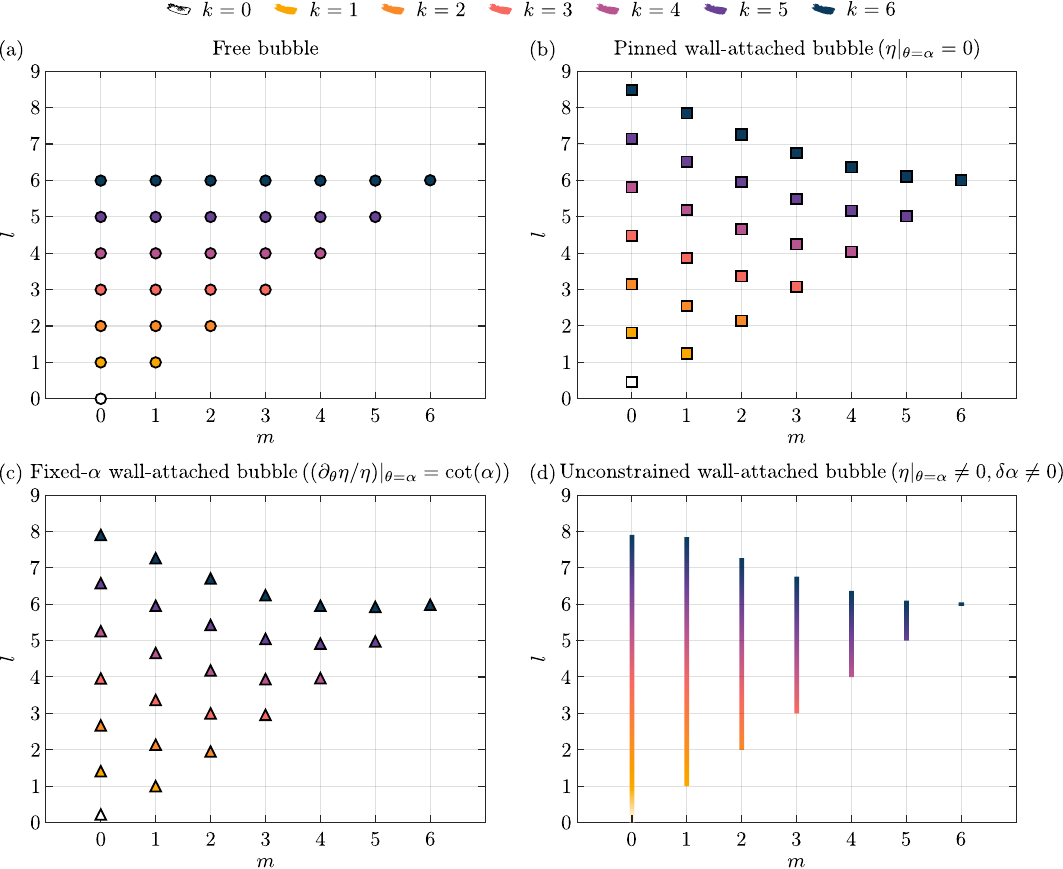}
    \caption{Spectrum of allowed shape modes for the first seven $k$ values, from 0 to 6, for, (a) a free bubble, (b) a pinned wall-attached bubble, (c) a fixed-contact-angle wall-attached bubble and, (d) an unconstrained wall-attached bubble. 
    The static contact angle considered for the wall-attached bubbles is $\alpha_0 = 3\pi/4$.
    $l$ and $m$ are the degree and order, respectively, of the spherical harmonic $Y_{l_{k}}^{m}$.
    $k$ is an index that sequentially orders shape modes of a specific order $m$. 
    For a free bubble, both $l$ and $m$ are integers, resulting in a degenerate spectrum where different shape modes share the same $l$.
    In contrast, for a pinned or fixed-contact-angle wall-attached bubble, $l$ is generally a non-integer, leading to a non-degenerate spectrum.
    For an unconstrained wall-attached bubble, the spectrum is no longer quantised in $l$ but is continuous and remains degenerate.
    }
    \label{fig:Spectrum}
\end{figure}

The tangential deformation $\tau(\theta,\phi,t)$ does not affect the curvature of the bubble, and thus, is irrelevant for analysing the dynamic behaviour of the interface.
To perform a spectral analysis on the normal deformation \(\eta(\theta, \phi, t)\), it is customary to expand it in a series of real spherical harmonics functions $Y_{l}^{m}(\theta,\phi)$ of degree $l$ and order $m$, with $0 \leq m \leq l$:
\begin{equation}
\eta(\theta,\phi,t) = \sum_m a_l^{m}(t) Y_{l}^{m}(\theta,\phi),
\end{equation}
where $a_l^{m}(t)$ represents the time-dependent amplitude of each spherical harmonic $Y_{l}^{m}(\theta,\phi) = N_{l}^{m} P_{l}^{m} (\cos (\theta)) \cos(m \phi)$.
$P_{l}^{m} (\cos (\theta))$ is a Legendre function and the normalisation factor $N_{l}^{m} = (\max(P_{l}^{m} (\cos (\theta)) \cos(m \phi)))^{-1}$ ensures that the spherical harmonic is constrained to a maximum magnitude of unity.
For a free bubble, which constitutes a closed surface, the following periodic boundary conditions must hold:
\begin{align}
&\eta(\theta,\phi,t) = \eta(\theta +2\pi,\phi,t), \quad \partial_{\theta}\eta(\theta,\phi,t) = \partial_{\theta}\eta(\theta +2\pi,\phi,t), \quad \forall \theta, \phi, t,\\
&\eta(\theta,\phi,t) = \eta(\theta,\phi +2\pi,t), \quad \partial_{\phi}\eta(\theta,\phi,t) = \partial_{\phi}\eta(\theta,\phi +2\pi,t), \quad \forall \theta, \phi, t.
\end{align}
This implies that the surface deformation can only be represented by Legendre functions with integer degrees $l$ and orders $m$, commonly known as associated Legendre polynomials.
From the linear framework used to derive equation (\ref{eq:DispRel}), it follows that the driving frequency and bubble radius determine the degree $l$ of the shape mode, but not its order $m$. Consequently, for a given $l$, there are $l+1$ possible values of $m$, each occurring with equal probability (see figure \ref{fig:ShapeModesIllustration}).
Modes that share the same resonance frequency, and thus the same $l$, are referred to as degenerate.
Nonlinear interactions dictate the combination of degenerate modes that ultimately emerges for each $l$.
The first seven modes with integer degree $l$ for $m=0$ are depicted in figures \ref{fig:FractionalLegendreFunction}(a)-(b).
The mode spectrum of a free bubble for the first seven degrees is depicted in figure \ref{fig:Spectrum}(a). 
The degree $l$ can be interpreted as the ``energy level" of the mode, while the order $m$ can be seen as the number of ``states" that become available with increasing $l$.
The ``energy level" $l$ can be augmented by increasing either the driving frequency or the bubble radius.

In general, spherical harmonics with an integer degree are not suitable for describing the surface deformation of a bubble in contact with a substrate, as they may not comply with the variational wetting condition (\ref{eq:vwr}).
For a pinned bubble, the deformation basis functions must satisfy the following boundary condition:
\begin{equation}\label{eq:LegEta0}
\eta|_{\theta=\alpha_0} = 0.
\end{equation}
This condition is fulfilled by Legendre functions $P_{l}^{m} (\cos (\theta))$ of real, though not necessarily integer, degree $l$ that are zero at $\theta=\alpha_0$.
Since the distinct real values of $l$ for which this condition holds depend separately on $m$, they will be denoted by $l_{k}(m)$, where $k$ is an integer index that sequentially orders the various roots $l$ at a given $m$.
The index $k$ is chosen to start at $m$, analogous to the case of integer $l$.
For a free bubble, $k$ corresponds to $l$.
In the azimuthal direction, periodic boundary conditions must hold:
\begin{align}
&\eta(\theta,\phi,t) = \eta(\theta,\phi +2\pi,t), \quad \partial_{\phi}\eta(\theta,\phi,t) = \partial_{\phi}\eta(\theta,\phi +2\pi,t), \quad \forall \theta, \phi, t,
\end{align}
which restrict the value of $m$ to an integer.
The first seven modes with real degrees $l$ for $m=0$ that comply with the condition (\ref{eq:LegEta0}) for $\alpha_0 = 3\pi/4$ are depicted in figure \ref{fig:FractionalLegendreFunction}(c)-(d).
The contact angle $\alpha_0 = 3\pi/4$ is chosen to align with the experimental observations discussed in the following sections.
The mode spectrum of a pinned bubble with $\alpha_0 = 3\pi/4$ for the first seven $k$ values is reported in table \ref{tab:kd}(a) and also depicted in figure \ref{fig:Spectrum}(b). 
In this case, shape modes are non-degenerate as each mode corresponds to an unique $l$ value.
This leads to a richer, more granular spectrum where all permissible modes can distinctly manifest.
It is important to recognise that a superposition of modes sharing the same degree $l$ but differing in order $m$ cannot satisfy the boundary condition along the entire contact line, since $\phi$-dependent deformation components cannot cancel out everywhere. Only individual modes that produce uniform deformation along the contact line can meet this constraint.
Compared to a free bubble, the $l$ value for a given $k$ value is higher, as the same number of shape mode features must be compressed into a shorter arc of the circumference.
Consequently, for a given radius of curvature, the value of $l$ increases as the contact angle decreases.
Finally, it can be observed that the deviation of $l$ from $k$ increases as $k$ increases and $m$ decreases.

If the contact line is free to move but the contact angle is fixed  ($\delta\alpha=0$), the deformation basis functions must satisfy the following boundary condition:
\begin{equation}\label{eq:LegDAlpha0}
({\partial_{\theta}\eta}/{\eta})|_{\theta=\alpha_0}  = \cot(\alpha_0).
\end{equation}
Again, this condition is met by Legendre functions $P_{l}^{m} (\cos (\theta))$ of real, though not necessarily integer, degree $l$ and integer order $m$.
The first seven modes with real degrees $l$ for $m=0$ that comply with the condition (\ref{eq:LegDAlpha0}) for $\alpha_0 = 3\pi/4$ are depicted in figure \ref{fig:FractionalLegendreFunction}(e)-(f).
The mode spectrum of a fixed-contact-angle bubble with $\alpha_0 = 3\pi/4$ for the first seven $k$ values is reported in table \ref{tab:kd}(b) and also depicted in figure \ref{fig:Spectrum}(c). 
The $l$ value for a given $k$ and $m$ values is lower compared to the pinned bubble case.
These findings indicate that the pinning conditions can markedly influence the resonance frequency of shape modes, including the breathing mode.

In the family of Legendre functions describing the deformation of a free bubble, characterised by integer indices $l$ and $m$, orthogonality is satisfied for both fixed $l$ and fixed $m$:
\begin{equation}
\left\langle  P_{l}^m, P_{l}^{m'}\right\rangle = \int_0^{\pi} \frac{P_{l}^m (\cos(\theta))  P_{l}^{m'} (\cos(\theta))}{1-\cos(\theta)^2} \sin(\theta) d\theta = 0, \quad \text{for $m \neq m'$}.
\end{equation}
\begin{equation}
\left\langle  P_{l}^m, P_{l'}^{m}\right\rangle = \int_0^{\pi} P_{l}^m (\cos(\theta))  P_{l'}^{m} (\cos(\theta)) \sin(\theta) d\theta = 0,  \quad \text{for $l \neq l'$}.
\end{equation}
The corresponding spherical harmonics $Y_{l}^m = N_{l}^m P_{l}^m (\cos(\theta)) \cos(m \phi) $ are thus orthogonal to each other:
\begin{equation}
\left\langle  Y_{l}^m, Y_{l'}^{m'}\right\rangle = \int_{\theta=0}^{\pi} \int_{\phi=0}^{2\pi}  Y_{l}^m Y_{l'}^{m'}  \sin(\theta) d\phi d\theta = 0,  \quad \text{for $l \neq l'$ and/or $m \neq m'$}.
\end{equation}
Orthogonality is also satisfied in the families of Legendre functions describing the deformation of a pinned bubble and a fixed-contact-angle bubble.
However, since no function shares the same $l$, orthogonality holds only for fixed $m$:
\begin{equation}
\left\langle  P_{l_k}^m, P_{l_{k'}}^{m}\right\rangle = \int_0^{\alpha_0} P_{l_k}^m (\cos(\theta))  P_{l_{k'}}^{m} (\cos(\theta)) \sin(\theta) d\theta = 0, \quad \text{for $k \neq k'$}.
\end{equation}
In these families as well, the corresponding spherical harmonics are therefore orthogonal to each other:
\begin{equation}
\left\langle  Y_{l_k}^m, Y_{l_{k'}}^{m'}\right\rangle = \int_{\theta=0}^{\alpha_0} \int_{\phi=0}^{2\pi}  Y_{l_k}^m Y_{l_{k'}}^{m'}  \sin(\theta) d\phi d\theta = 0,  \quad \text{for $k \neq k'$ and/or $m \neq m'$}.
\end{equation}
Spherical harmonics belonging to different families are in general not orthogonal, and the inner product between them is non-zero.
Consequently, the spherical harmonics in these two families are not volume-preserving as they are not orthogonal to $Y_0^0=1$, and therefore their 1-norm is non-zero.

If both the contact line and contact angle are free to move, the deformation can adopt the shape of a Legendre function with any real $l$ value.
This is because there are no boundary conditions imposed along the $\theta$-direction, including the periodicity condition required for a free bubble.
Consequently, as shown in figure \ref{fig:Spectrum}(d), the mode spectrum for a bubble with a free contact line and contact angle is no longer quantised in $l$, but rather, it becomes continuous. 
However, it remains quantised in $m$ due to the $2\pi$-periodicity in the $\phi$-direction that still holds.
In this case, the spectrum exhibits degeneracy, as shape modes can share the same degree $l$, a scenario analogous to that of a free bubble.

\section{Experimental observation of the bubble response}\label{Sec4}
\begin{figure}
    \centering   \includegraphics[width=0.95\columnwidth]{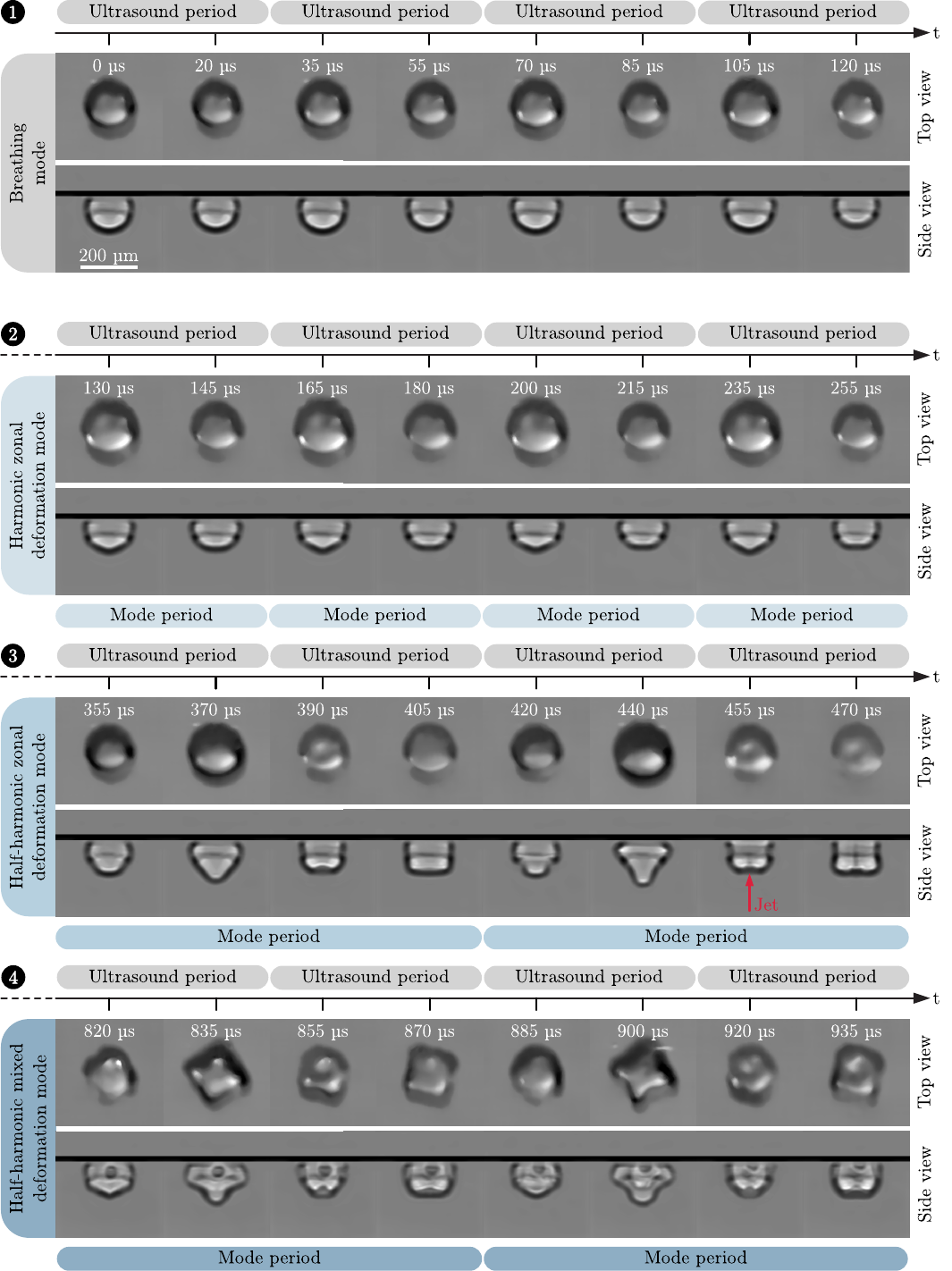}
    \caption{Overview of the response of an air bubble in contact with a glass substrate and subjected to ultrasound.
    The equilibrium radius of the bubble is $R_0 = \SI{82}{\micro\meter}$ and the ultrasound frequency is $f_{\rm d} = \SI{30}{\kilo\hertz}$, while its peak pressure is $p_{\rm a} = \SI{2.7}{\kilo\pascal}$.
    The response can be categorised into four sequential regimes: (i) breathing mode, (ii) harmonic zonal deformation mode, (iii) half-harmonic zonal deformation mode, and (iv) combined half-harmonic zonal and sectoral deformation mode.
    The top rows display the bubble dynamics from a top-view perspective captured with visible light, while the bottom rows present a side-view perspective, captured using X-ray synchrotron radiation.
    When the amplitude of the deformation modes exceeds a certain threshold, jetting can occur.
    The images have been denoised and their backgrounds have been removed for visual clarity.
    }
    \label{fig:TimelineCollage}
\end{figure}
We present a representative example of the response of an air bubble resting against the glass substrate subjected to ultrasound driving in figure \ref{fig:TimelineCollage} (and supplementary movie 1).
In the figure, the top rows offer a top-view perspective captured with visible light, while the bottom rows show a side-view obtained using X-ray synchrotron light.
The frames have been post-processed to enhance visual clarity: noise has been reduced, and the static background has been subtracted.
In the top-view images, slight shadowing effects are visible due to the angled illumination employed during capture.
In the side-view images, a faint horizontal line appears as an X‑ray imaging artefact, most likely resulting from the substrate not being perfectly parallel to the X-ray beam.
This artefact is not related to the physical structure or properties of the bubble and does not influence the interpretation of the data presented.
The equilibrium radius of the bubble is $R_0 = \SI{82}{\micro\meter}$ and the equilibrium contact angle is $\alpha_0 \approx 140^\circ$, as directly estimated from the images.
This measured contact angle is consistent with values reported in previous experimental studies involving air bubbles in water in contact with untreated glass surfaces \citep{Ozkan2017InterpretingSurfaces}.
The ultrasound frequency is $f_{\rm d} = \SI{30}{\kilo\hertz}$, its peak pressure is $p_{\rm a} = \SI{2.7}{\kilo\pascal}$, and its direction in the images is from bottom to top. 
The wavelength $\lambda$ of the acoustic forcing is considerably larger than the bubble size: $\lambda/2R_0 \sim O(10^2)$. 
Therefore, the pressure field can be considered uniform around the bubble.
Here, and in all subsequent figures, the time $t = 0$ marks the onset of ultrasound excitation.

The time evolution of the bubble response can be categorised into four distinct phases, as elaborated below and in the following section.
In the initial acoustic cycles, the bubble response exclusively exhibits a breathing mode, as illustrated in the first segment of the figure.
During this phase, the bubble undergoes compression and expansion cycles at the frequency of the ultrasound driving while maintaining its spherical shape.
As time proceeds, the bubble tends to flatten slightly against the substrate due to the primary radiation force exerted by the ultrasound.
Moreover, if the breathing mode exceeds a certain threshold in amplitude, the bubble interface destabilises leading to the development of a standing wave pattern.
Initially, this pattern is axisymmetric, its amplitude small, and it matches the frequency of the acoustic driving, as outlined in the second segment of the figure. 
Therefore, following the nomenclature for spherical harmonics, this initial pattern is zonal and its frequency classifies it as harmonic.
Then, the deformation pattern lowers its frequency to half the acoustic driving frequency and its amplitude increases, which is depicted in the third segment of the figure.
The zonal standing wave pattern is thus now half-harmonic, which is a signature trait of the Faraday instability.
Interestingly, owing to the more intense shape deformation following the frequency shift, the bubble tip undergoes a vigorous folding when the shape mode reverses, ultimately resulting in the formation of a microjet that impacts the substrate, as highlighted by the red arrow in the corresponding frame.
The occurrence of jetting is cyclical and does not result in bubble disruption.
As the microjet manifestation is linked to the shape mode cyclicity, its frequency is half-harmonic.
The jet retraction may result in the pinching-off of one or more droplets due to the Rayleigh-Plateau instability.
Over time, the bubble begins to show azimuthal perturbations that break its axial symmetry and lead to the emergence of a sectoral standing wave pattern.
The zonal pattern persists but diminishes in intensity, leading to the cessation of jetting as the tip folding weakens compared to before.
The mixing of a zonal pattern and a sectoral pattern marks the ultimate phase of bubble response, as reported in the fourth segment of the figure.
Before the jetting stops, the bubble under examination exhibits six consecutive jets.
Finally, the images further reveal that, for the glass substrate being studied, both the movement of the contact line and changes in the contact angle are allowed as the bubble interface evolves.
This behaviour aligns more closely with the fully-free dynamical wetting conditions occurring in the ideal case of an unconstrained wall-attached bubble, as depicted in figure \ref{fig:Spectrum}(d).

\section{Detailed analysis of the bubble response}\label{Sec5}
\subsection{Breathing mode}\label{BrMode}
The bubble deformation $\eta(\theta,\phi, t)$ can be decomposed into a volume-variant component $\eta_{\rm volume}(\theta,\phi, t)$ and a volume-invariant component $\eta_{\rm shape}(\theta,\phi, t)$:
\begin{equation}
    \eta(\theta,\phi, t) = \eta_{\rm volume}(\theta,\phi, t) +  \eta_{\rm shape}(\theta,\phi, t).
\end{equation}
For a free bubble,  $\eta_{\rm volume}(\theta,\phi, t)=a_0^0(t)Y_0^0(\theta,\phi)$, which corresponds to the breathing mode.
The shape deformation, on the other hand, is described by $\eta_{\rm shape}(\theta,\phi, t)=\sum a_l^m(t) Y_l^m(\theta,\phi)$ with $l\neq0$.
For a wall-attached bubble, performing such a decomposition in terms of spherical harmonics is not straightforward.
As mentioned previously, there are no spherical harmonics that simultaneously satisfy the wetting condition and are volume invariant.
Nevertheless, we can still experimentally track the temporal evolution of the volumetric change, define an equivalent breathing mode amplitude, and compare it with classical theoretical models for spherical bubbles.
This proves useful to elucidate the accuracy of these models in the context of wall-attached bubbles and serves as a valuable means to validate the value of the acoustic driving pressure measured by the hydrophone.
For this purpose, we define an average deformation of the bubble surface, analogous to an increase $\delta R(t)$ in the radius of the spherical cap,
$
    \delta R(t) = \frac{1}{2\pi(1-\cos(\alpha_0))}\int_0^{2\pi} \int_0^{\alpha_0} \eta(\theta,\phi,t)\sin{\theta}d\theta d\phi.
$
The volume of this equivalent spherical cap is consequently given by
$
V(t) = \left(\frac{2}{3} - \cos(\alpha_0) + \frac{1}{3}\cos^3(\alpha_0)\right)\pi\left(R_0 + \delta R(t)\right)^3.
$
The radius $R_{\rm eq}(t)$ of a spherical bubble, which has the same volume as the equivalent spherical cap, is given by
$
R_{\rm eq}(t)  =  \left (\frac{3V(t)}{4 \pi} \right ) ^{{1}/{3}} = \left(\frac{1}{2} -\frac{3}{4}\cos(\alpha_0) + \frac{1}{4}\cos^3(\alpha_0)\right)^{{1}/{3}}\left(R_0+ \delta R(t) \right).
$
For a contact angle $\alpha_0 \approx 140^\circ$, $R_{\rm eq}(t) \approx 0.99 \left(R_0+ \delta R(t) \right)$.
Therefore, from now on, the difference between the two will be regarded as negligible, and the term $R(t)$ will be used.

To experimentally measure $R(t)$ we employ an image-processing algorithm to extract the bubble contour from the side-view video recordings.
As long as the bubble maintains axisymmetry, i.e., before developing sectoral shape modes, its volume $V(t)$ can be determined by integrating the area of disks stacked along the axis of symmetry $z$, defined by the bubble contour $R(z,t)$, as
$
V(t) =\pi \int_{z^{-}}^{z^{+}}  R(z,t)^2 dz,
$
where $z^{-}$ and $z^{+}$ represent the limiting values within which the contour of the bubble $R(z,t)$ is defined.
The radius $R(t)$ of a volume-equivalent spherical bubble can then be determined as:
\begin{equation}
R(t)  =  \left (\frac{3V(t)}{4 \pi} \right ) ^{1/3}.
\end{equation}

\begin{figure} 
    \centering 
    \includegraphics[width=\columnwidth]{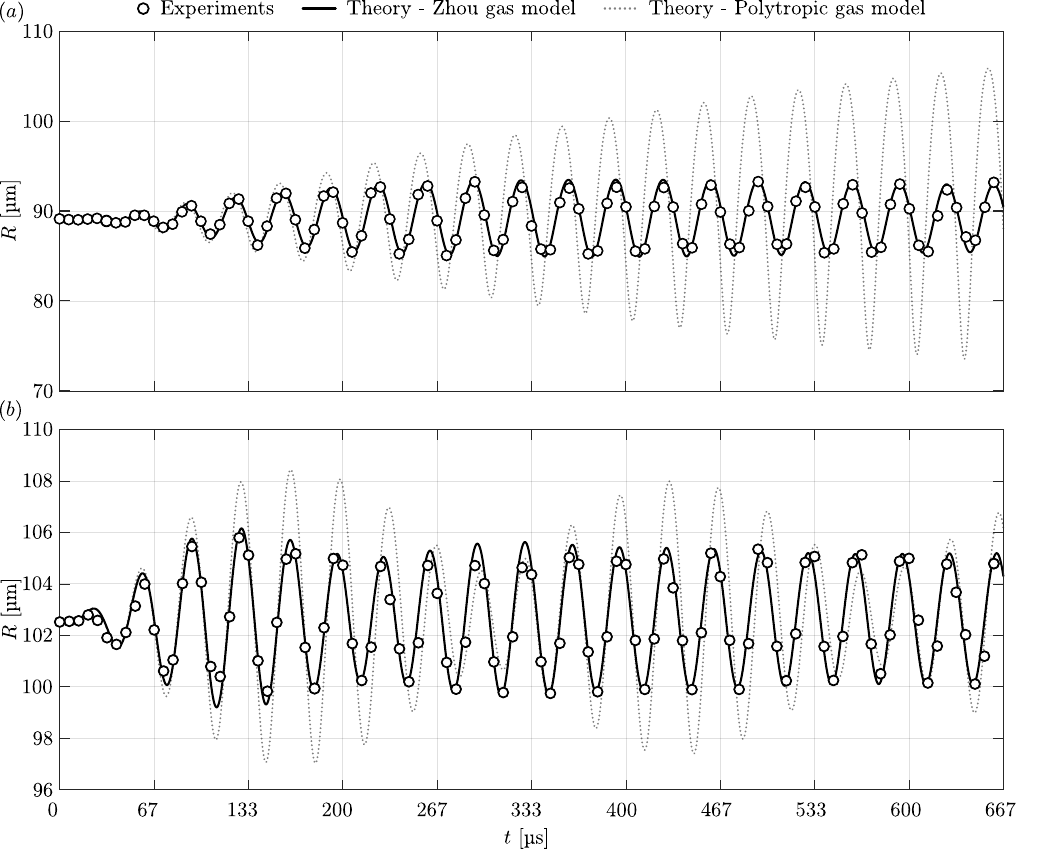}
    \caption{Experimental and predicted time evolution of the equivalent radius of a bubble in contact with a glass substrate subjected to an ultrasound driving at a frequency $f_{\rm d} = \SI{30}{\kilo\hertz}$.
    In (a), the bubble has an an equilibrium radius $R_0 = \SI{89.0}{\micro\meter}$ and the ultrasound pressure amplitude is $p_{\rm a} = \SI{2.1}{\kilo\pascal}$.
    In (b), the bubble has an an equilibrium radius $R_0 = \SI{102.6}{\micro\meter}$ and the ultrasound pressure amplitude is $p_{\rm a} = \SI{4.6}{\kilo\pascal}$.
    Theoretical predictions are obtained by solving the Rayleigh-Plesset equation using the following physical parameters: $p_{\infty} = \SI{101.0}{\kilo\pascal}$, $\sigma_{\rm lg} = \SI{72}{\milli\newton\per\meter}$, $c_{\rm l} = \SI{1481}{\meter\per\second}$, $\mu_{\rm l} = \SI{9.54e-4}{\pascal\second}$, and $\rho_{\rm l} = \SI{997.8}{\kilo\gram\per\cubic\meter}$. 
    The prediction obtained using the polytropic gas model, with $n = \gamma = 1.4$, is shown by the dotted gray line.
    The prediction obtained using the Zhou gas model, with parameters $\gamma=1.4$, $K_{\rm g} = \SI{0.026}{\watt\per\meter\per\kelvin}$, $\mathcal{R} = \SI{287}{\joule\per\kilo\gram\per\kelvin}$, and $T_{\rm g}|_{R} = \SI{295}{\kelvin}$, is represented by the continuous black line.}
    \label{fig:RtCurve}
\end{figure}

The radial motion of this equivalent spherical bubble of radius $R(t)$ in proximity to a solid boundary and subjected to a spatially uniform acoustic field, can be approximated by the Rayleigh–Plesset equation for mildly compressible Newtonian fluids.
The influence of the nearby solid boundary is incorporated using the method of images \citep{Strasberg1953TheLiquids}, yielding the following formulation:
\begin{equation}\label{eq:RPplusImage}
\rho_{\rm l} \left(R \ddot R + \frac{3}{2} \dot R^2 \right)= \left( 1 + \frac{R}{c_{\rm l}} \frac{d}{dt} \right) p_{\rm g}
   -2\frac{\sigma_{\rm lg}}{R} - p_{\infty} - p_{\rm d}\left(t\right) - \rho_{\rm l} \frac{2R\dot{R}^2 + R^2 \ddot{R}}{2d} - 4\mu_{\rm l} \frac{\dot R}{R},
\end{equation}
where over-dots denote time differentiation, $\rho_{\rm l}$ the liquid density, $c_{\rm l}$ the speed of sound in the medium, $p_{\rm g}$ the gas pressure within the bubble, $\sigma_{\rm lg}$ the liquid-gas surface tension,  $p_{\infty}$ the undisturbed ambient pressure, $p_{\rm d}(t)$ the acoustic driving pressure, $\mu_{\rm l}$ the dynamic viscosity of the medium and  $d=R_0$ is the distance from the centre of the bubble to the wall.
A full derivation of the pressure term accounting for the rigid boundary is provided in Appendix \ref{appC}.
The method of images is typically valid when the distance between the bubble and the wall is much larger than the bubble radius. In our case, where the bubble is attached to the wall, finite-size and multipole effects may become significant. 
More advanced models have been developed to describe bubble dynamics near boundaries \citep{Hay2013ModelLayers,Doinikov2011AcousticThickness}, which capture more complex interactions than the simple image method.
Nevertheless, despite its simplicity, the method of images accurately predicts the experimentally observed resonance size and damping of wall-attached bubbles, showing strong agreement with both radial dynamics (figure~\ref{fig:RtCurve}) and shape mode pressure inception (figure~\ref{fig:R0vPa}), which we will examine in detail later.
To solve equation~\ref{eq:RPplusImage}, it is necessary to determine the gas pressure $p_{\rm g}$ within the bubble. 
Given its simplicity, the assumption of a polytropic process is widely employed:
$
p_{\rm g} = p_{\rm g,0} \left(\frac{R_0}{R}\right)^{3n}, \ p_{\rm g,0} = p_{\infty} + \frac{2\sigma_{\rm lg}}{R_0},
$
with $p_{\rm g,0}$ denoting the bubble internal pressure at rest, $R_0$ the equilibrium bubble radius, and $n$ the polytropic index.

In figure \ref{fig:RtCurve}, the predictions of the theoretical model are compared with the experimental radial time evolution of bubbles subjected to ultrasound forcing at a frequency of $f_{\rm d} = \SI{30}{\kilo\hertz}$.
We assume for the polytropic gas model that the gas within the bubble undergoes an adiabatic process, characterised by an exponent $n = \gamma$, where $\gamma = 1.4$ denotes the specific heat ratio for the gas. 
This assumption is justified by the high Péclet number, $ \text{Pe} = {R_0^2 \omega_{\rm d}}/{D_{\rm g}} \sim 100$, where $\omega_{\rm d}$ is the driving angular frequency and $D_{\rm g}$ represents the thermal diffusivity of the gas.
Subfigure \ref{fig:RtCurve}(a) illustrates the case of a resonant-sized bubble with an equilibrium radius $R_0 = \SI{89.0}{\micro\meter}$ and ultrasound pressure amplitude $p_{\rm a} = \SI{2.1}{\kilo\pascal}$.
Here, the theoretical prediction based on the polytropic gas model   significantly overestimates the experimental results and manifests a continuous growth behaviour. 
Conversely, subfigure \ref{fig:RtCurve}(b) shows results for an over-resonant bubble with an equilibrium radius $R_0 = \SI{102.6}{\micro\meter}$ and ultrasound pressure amplitude $p_{\rm a} = \SI{4.6}{\kilo\pascal}$.
In this case, the theoretical curve obtained using the polytropic gas model exhibits a pronounced beating pattern, which is absent in the experimental data.
A beating phenomenon arises from the interference between two signals with different frequencies, $f_1$ and $f_2$, where the beat frequency is given by $f_{\rm beat} = |f_2 - f_1|$.
The observed beat frequency of approximately $\SI{4}{\kilo\hertz}$ suggests that the second signal, which overlaps with the forced response of the bubble at $\SI{30}{\kilo\hertz}$, corresponds to the free response at the resonance frequency, which for this bubble size is approximately $\SI{26}{\kilo\hertz}$.
From this, we deduce that the strong overestimation seen in subfigure \ref{fig:RtCurve}(a) results from constructive interference between the forced and free responses, both occurring at approximately the same frequency for this bubble size.
We attribute the anomalous, underdamped free response, absent in the experimental observations, to the lack of thermal damping in the polytropic gas model.
Thermal damping, as discussed by \citet{Chapman1971ThermalBubbles}, is the dominant damping mechanism for bubbles of the sizes considered in this study.

To address the thermal interaction, we employ the Zhou model \citep{Zhou2021ModelingBubble} for the bubble gas pressure $p_{\rm g}$.
A detailed description of the model is provided in Appendix \ref{appD}.
Figure \ref{fig:RtCurve} illustrates that the theoretical predictions derived from this model align remarkably well with the experimental results, showing a much milder beating pattern.

\subsection{Meniscus waves}
\begin{figure} 
    \centering
        \includegraphics[width=\columnwidth]{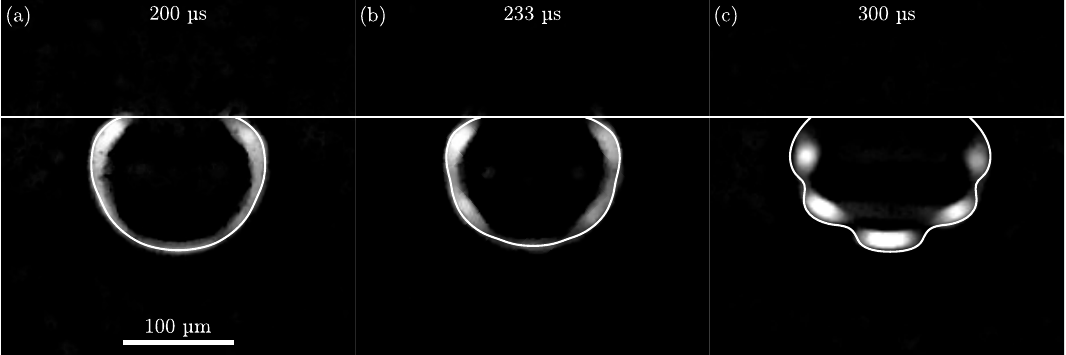}
    \caption{Time evolution of harmonic meniscus waves produced by the harmonic acceleration of the bubble surface close to the wall.
    The wave pattern is highlighted by performing a subtraction operation between the X-ray image capturing the wave at its peak and the one with the wave at its trough.
    (a) Initial emission of the travelling meniscus wave. 
    (b) Propagation of the travelling meniscus wave. 
    (c) Steady-state of meniscus waves forming a standing wave pattern.}
    \label{fig:MeniscusWave}
\end{figure}

Upon bubble oscillations, travelling surface waves are emitted from the pinning line. 
These waves move towards the bubble symmetry axis, $z$, and equilibrate to form an axisymmetric harmonic standing wave that oscillates at the same frequency as the ultrasound applied.
Figure \ref{fig:MeniscusWave} displays the initial emission of the wave, its propagation, and the formation of a steady-state pattern.
This phenomenon has previously been observed and investigated on flat liquid surfaces within containers, where a meniscus forms along the walls \citep{Douady1990ExperimentalInstability,Torres1995Five-foldExperiment,Batson2013TheNon-ideality,Ward2019FaradayLayers,Shao2021SurfaceContainer}.
When the container is subjected to harmonic vertical acceleration, travelling edge waves are emitted from the walls.
These waves are not generated by a parametric instability but rather by synchronous adjustments of the meniscus profile in response to the harmonic modulation of the acceleration field, resulting in the harmonic emission of an axisymmetric wave from the wall in order to conserve fluid mass.
Their occurrence requires no minimum threshold in the magnitude of the driving acceleration, and their amplitude is linearly proportional to this magnitude.
The meniscus forms when the contact angle deviates from $90\degree$, and meniscus waves occur regardless of whether the contact line is pinned or free.
These waves can be suppressed by filling the liquid container to the brim, thereby pinning the contact line at the brim with a contact angle of $90\degree$.
To our knowledge, our study is the first to recognise the presence of meniscus waves on ultrasound-driven bubbles in contact with a solid surface.
In this case, the presence of a wall mandates that the acceleration of the bubble interface near the wall must be parallel to the wall, as dictated by the non-penetration condition.
As a result, when the contact angle deviates from $90\degree$, the acceleration along the pinning line of the bubble is not orthogonal to the bubble surface and meniscus waves are emitted.
Conversely, when a bubble in contact with a wall exhibits a contact angle of $90\degree$ meniscus waves are not expected to occur, as the direction of acceleration is orthogonal to the bubble surface at all points.

\subsection{Faraday waves}

\begin{figure} 
    \centering
        \includegraphics[width=\columnwidth]{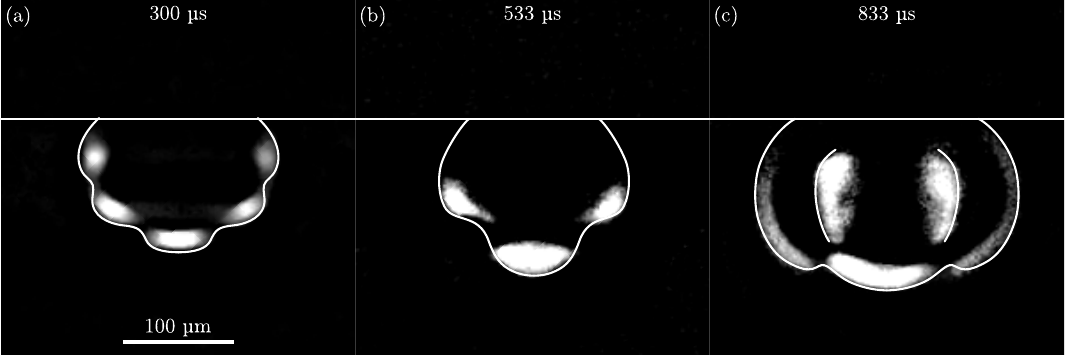}
    \caption{Comparison between meniscus waves and Faraday waves on the same bubble of figure~\ref{fig:MeniscusWave}.
        (a) Meniscus waves are harmonic waves whose emergence is not contingent upon surpassing a minimum amplitude threshold in the ultrasound driving.
        (b) Faraday waves are half-harmonic waves characterised by a wavelength approximately double that of meniscus waves and a significantly higher amplitude.
        Their occurrence is contingent upon surpassing a critical threshold in the ultrasound driving amplitude.   
        Initially, Faraday waves are axisymmetric, i.e. the shape mode is purely zonal.
        (c) Over time, Faraday waves gradually lose their axisymmetry, manifesting a sectoral mode that overlays the pre-existing zonal mode.
        This transition is marked by the emergence of lobes in both the polar and azimuthal directions.
        The persistence of the bottom lobe indicates the absence of tesseral modes.
        }
    \label{fig:Meniscus_v_Faraday}
\end{figure}

When the amplitude of the ultrasound pressure exceeds a critical threshold, a half-harmonic parametric instability known as Faraday instability emerges, supplanting the harmonic pattern of meniscus waves.
The resulting standing wave pattern, also termed shape mode, is the result of  the nonlinear interaction between the external periodic forcing and the natural frequencies of the bubble interface.
This interaction leads to the amplification of certain modes while damping others, resulting in the observable patterns.
Initially, the Faraday wave pattern exhibits axisymmetry, akin to meniscus waves, resulting in a purely zonal shape mode.
The wavelength of Faraday waves is approximately twice that of meniscus waves, and its amplitude is substantially higher, as evidenced in figure \ref{fig:Meniscus_v_Faraday}(a)-(b).
As time proceeds, Faraday waves progressively lose their axisymmetry, ultimately stabilising into a non-axisymmetric shape mode.
This finding supports the theoretical prediction for spherical entities as stated by \cite{Chossat1991Steady-State03-Symmetry}, which asserts that no axisymmetric mode with $l \geq 2$ is inherently stable.
The ultimate shape mode of the bubble results from the superposition of a sectoral mode onto the pre-existing zonal mode, as depicted in figure \ref{fig:Meniscus_v_Faraday}(c).
This transition is evidenced by the development of lobes in the azimuthal direction, which overlap with the existing axisymmetric lobes in the polar direction.
The absence of layer-symmetrical patterns indicates that tesseral shape modes are not present.
This contrasts with the findings of \cite{Fauconnier2020NonsphericalBubble}, who reported the presence of tesseral modes.
However, their study employed a different experimental setup: the substrate was made of PMMA, the ultrasound was oriented parallel to the substrate, and shape modes were identified solely from a top-view perspective.

\begin{figure} 
    \centering
        \includegraphics[width=\columnwidth]{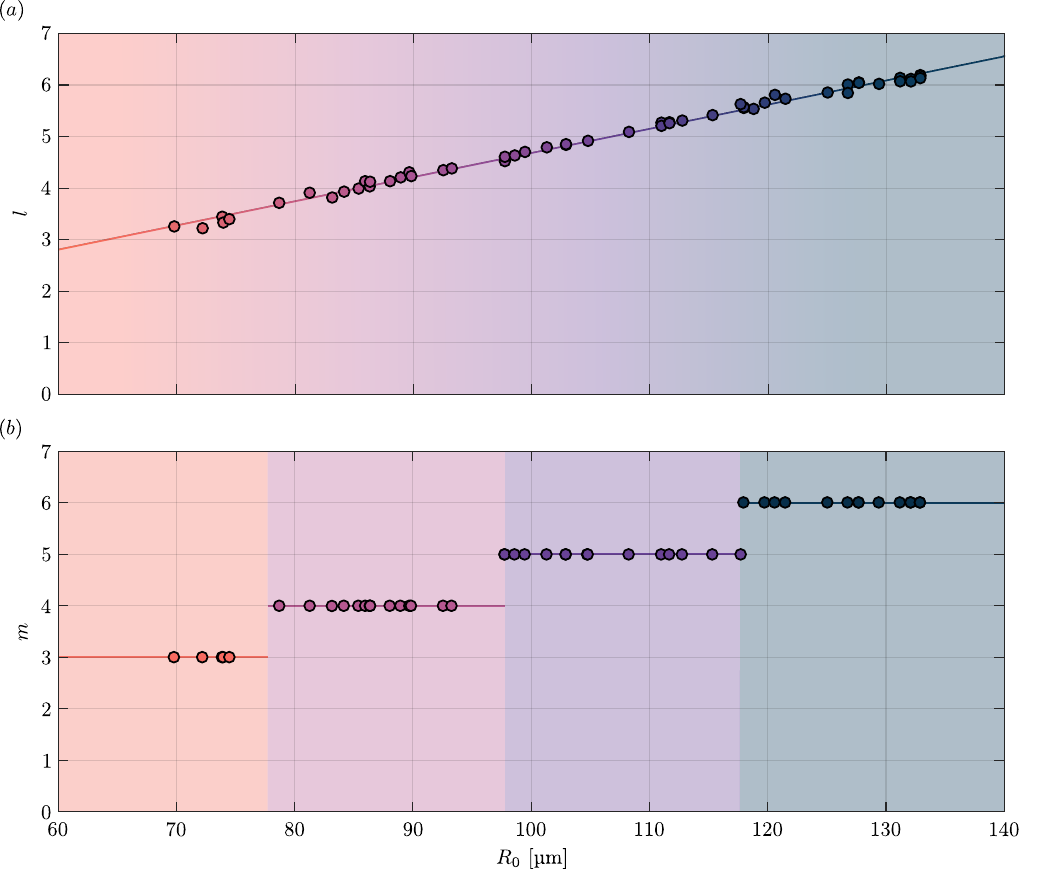}
    \caption{Experimentally measured shape mode degree (a) and order (b) of bubbles in contact with a glass substrate, subjected to an ultrasound driving frequency of $f_{\rm d} = \SI{30}{\kilo\hertz}$, as a function of their equilibrium radius.
    The degree varies continuously with the bubble radius, whereas the order is quantised and only assumes integer values.}
    \label{fig:Exp_l&m}
\end{figure}
Given the truncated geometry of a wall-attached bubble, the degree $l$ of the observed zonal shape mode, generally a real number, can be estimated only by measuring the spacing between the ridges of the shape mode as seen from the side view and comparing it with the circumference of the bubble.
The length of the arc $a$ subtended by the chord of length $c$ joining two ridges of the shape mode is calculated as:
\begin{equation}
a = 2R_0\sin^{-1}\left(\frac{c}{2 R_0}\right).
\end{equation}
Thus, $l$ is given by the number of times the arc length can divide the bubble circumference:
\begin{equation}
l = 2\pi \frac{R_0}{a}.
\end{equation}
Conversely, the order $m$ of the observed sectoral shape mode can be readily determined by counting the number of lobes along the azimuthal direction when the bubble is viewed from above.
Since the bubble contour remains closed and periodic in this view, $m$ must be an integer.
At a fixed ultrasound driving frequency $f_{\rm d}=\SI{30}{\kilo\hertz}$, the measured values of $l$ vary approximately linearly with the bubble equilibrium radius $R_0$, while the values of $m$ show a stepwise progression, as illustrated in figure~\ref{fig:Exp_l&m}.
The fact that $l$ spans a practically continuous range—unlike the quantised behaviour expected under strict boundary conditions—suggests that the contour endpoints of the bubble are effectively free. This interpretation is supported by the observed motion of the contact line and changes in contact angle during bubble oscillation. Thus, the current configuration can be considered corresponding to the idealised scenario of a wall-attached bubble with free boundary conditions, exhibiting a spectrum of shape modes that is continuous in $l$ and discrete in $m$, as discussed in section~\ref{Sec32} and shown in figure~\ref{fig:Spectrum}(d).
It can be determined from figure \ref{fig:Exp_l&m} that, for a given radius $R_0$, the value of $m$ of the sectoral mode that manifests is approximately the integer closest to the observed value of $l$ for that radius.
Therefore, the bubble deformation $\eta(\theta,\phi,t)$ can be decomposed as:
\begin{equation}
    \eta(\theta,\phi,t) = \underbrace{a_0^0(t)Y_0^0(\theta,\phi)}_\text{Breathing mode} + \underbrace{a_l^0(t)Y_l^0(\theta,\phi)}_\text{Zonal mode} + \underbrace{a_l^{\lfloor l \rceil}(t)Y_l^{\lfloor l \rceil}(\theta,\phi)}_\text{Sectoral mode},
\end{equation}
where $\lfloor \rceil$ indicates the rounding operator.
It is important to note that, for each value of $l$, only a specific pair of values of $m$ manifest from the set of possible values, specifically the zeroth value and the highest possible value of $m$ for the given $l$.
Consequently, under the present wetting conditions, the spectrum of shape modes is degenerate, similar to that of a free bubble \citep{Cattaneo2025CyclicDelivery}, with the difference that the values of $l$ are real and continuous rather than integer.

\begin{figure}
    \centering
        \includegraphics[width=\columnwidth]{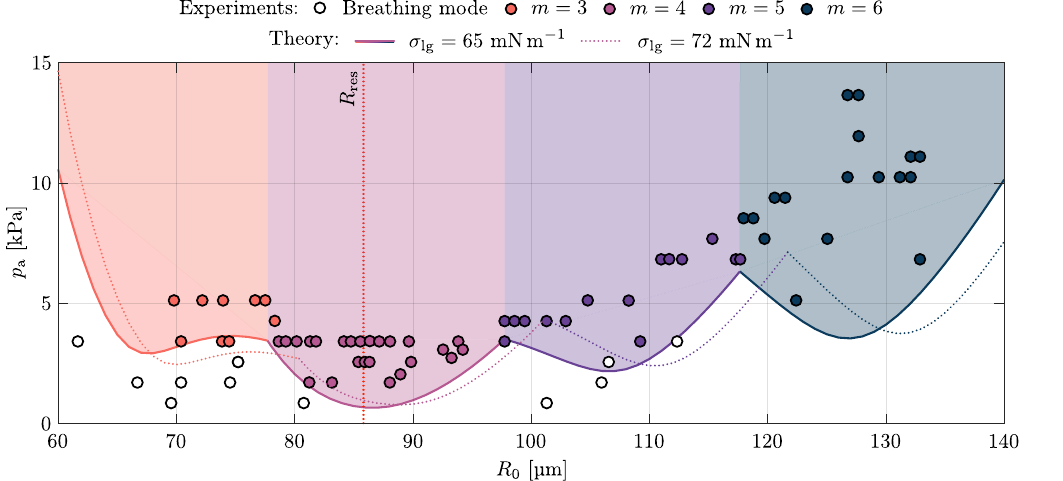}
    \caption{Shape mode order $m$ of bubbles in contact with a glass substrate, subjected to an ultrasound driving frequency $f_{\rm d} = \SI{30}{\kilo\hertz}$, as function of the equilibrium radius and the ultrasound pressure amplitude. 
    Experimental data are depicted with symbols, while the theoretical threshold for the onset of shape modes is based on the model from \cite{Francescutto1978PulsationBubbles}, as detailed in the text.
    The solid lines correspond to a surface tension value $\sigma_{\rm lg} = \SI{65}{\milli\newton\per\meter}$, and the dotted lines represent $\sigma_{\rm lg} = \SI{72}{\milli\newton\per\meter}$.
    The other parameters used in the model are: $\rho_{\rm l} = \SI{997.8}{\kilo\gram\per\cubic\meter}$, $\mu_{\rm l} = \SI{9.54e-4}{\pascal\second}$ and $\mu_{\rm e}$ extracted from the theoretical plots in \cite{Chapman1971ThermalBubbles}.
    }
    \label{fig:R0vPa}
\end{figure}
The occurrence of the Faraday instability, as detected experimentally in relation to bubble size and ultrasound pressure amplitude, is summarised in figure \ref{fig:R0vPa}.
The instability region appears to be arranged into distinct tongues, each corresponding to the specific order $m$ observed.
This finding is notable as the presence of distinct tongues indicates that the shape modes exhibit discrete resonance sizes for a fixed acoustic driving frequency.
Given that the accessible range of shape mode degrees $l$ is continuous, one would expect the associated resonant bubble sizes to be similarly continuous.
However, the quantisation of the azimuthal wavenumber $m$ introduces a selectivity on the bubble size, resulting in a discrete number of resonance sizes.
The experimentally observed instability thresholds agree well with those predicted by the theory from \cite{Francescutto1978PulsationBubbles} and \cite{Nabergoj1978OnBubbles} developed for shape oscillations of free bubbles.
Full details of our implementation of the Francescutto \& Nabergoj model for wall-attached bubbles are provided in Appendix \ref{appE}.
The minimum inception threshold for shape modes occurs at the bubble radius that corresponds to the resonant radius, which is also close to the resonance size for the shape mode with $m=4$, making it the most easily excited shape mode.
We note that to achieve alignment with the experimental data, we found necessary to use a surface tension value of $\SI{65}{\milli\newton\per\meter}$.
This is lower than the surface tension of a clean air-water interface, which is around $\SI{72}{\milli\newton\per\meter}$.
The reduction in surface tension could be attributed to the presence of surfactants in the liquid medium that have adsorbed onto the bubble interface and thus decreased its surface tension.
This value closely aligns with the one reported in \cite{Versluis2010MicrobubbleDriving}, where the authors investigated the shape modes of free air bubbles in an aqueous medium driven by ultrasound.
The theory predicted by \cite{Francescutto1978PulsationBubbles} achieves strong agreement with experiments even without accounting for meniscus waves, suggesting that these waves are unlikely to be a significant factor in triggering shape modes.
This observation aligns with the findings of \cite{Batson2013TheNon-ideality}, who determined in the context of flat water-air interfaces that significant influence of meniscus waves is more likely to occur for harmonic axisymmetric Faraday waves, as they share the same frequency and pattern with meniscus waves.

\subsection{Time evolution of shape modes}
\begin{figure} 
    \centering
        \includegraphics[width=\columnwidth]{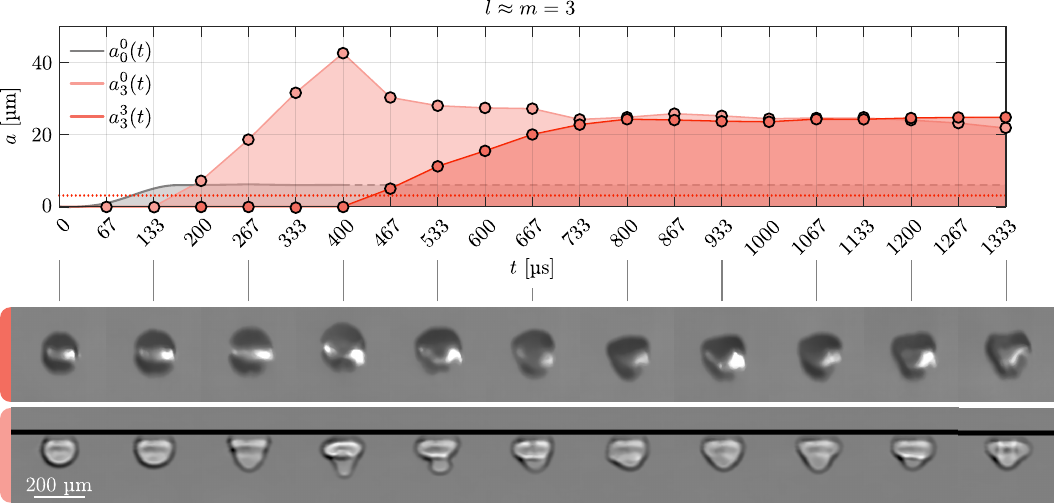}
    \caption{Time evolution of the experimental amplitude of the breathing, zonal and sectoral mode for the case $l \approx m=3$ ($l =3.22$)}.
    The red dotted line marks the threshold amplitude of the breathing mode for the onset of shape modes.
    The amplitude of the breathing mode is shown as dashed during the occurrence of the sectoral mode, indicating that accurate measurement is not possible due to the loss of bubble axisymmetry.
    Below is a sequence of top-down optical and side-view X-ray images illustrating the dynamics of the bubble.
    The images have been denoised and their backgrounds have been removed for visual clarity.
    The equilibrium bubble radius is $R_0 = \SI{72} {\micro\meter}$.
    The ultrasound driving frequency is $f_{\rm d} = \SI{30}{\kilo\hertz}$ and its amplitude is $p_{\rm a} = \SI{5.3} {\kilo\pascal}$.
    \label{fig:AnalysisN3}
\end{figure}

\begin{figure} 
    \centering
        \includegraphics[width=\columnwidth]{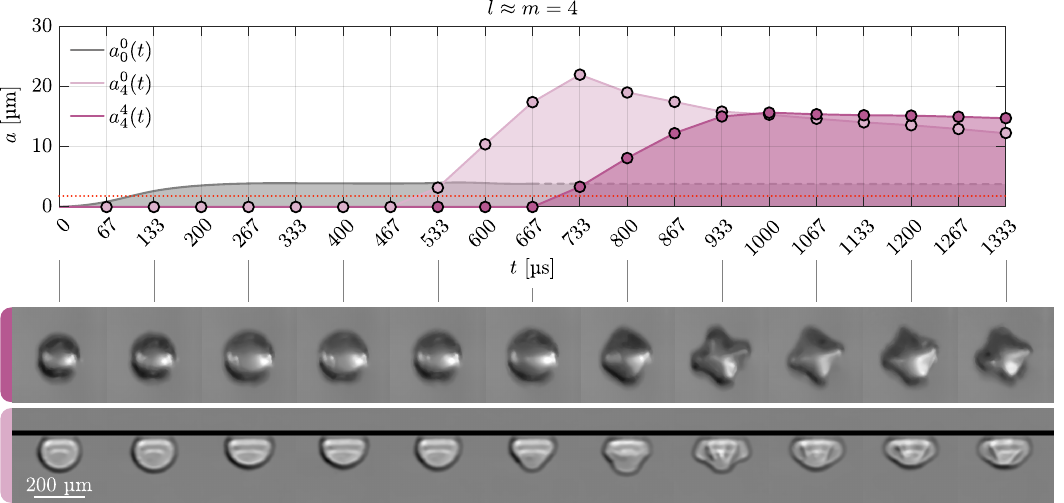}
    \caption{Time evolution of the experimental amplitude of the breathing, zonal and sectoral mode for the case $l \approx m=4$ ($l = 4.21$)}.
    The red dotted line marks the threshold amplitude of the breathing mode for the onset of shape modes.
    The amplitude of the breathing mode is shown as dashed during the occurrence of the sectoral mode, indicating that accurate measurement is not possible due to the loss of bubble axisymmetry.
    Below is a sequence of top-down optical and side-view X-ray images illustrating the dynamics of the bubble.
    The images have been denoised and their backgrounds have been removed for visual clarity.
    The equilibrium bubble radius is $R_0 = \SI{89} {\micro\meter}$.
    The ultrasound driving frequency is $f_{\rm d} = \SI{30}{\kilo\hertz}$ and its amplitude is $p_{\rm a} = \SI{2.1} {\kilo\pascal}$.
    \label{fig:AnalysisN4}
\end{figure}

\begin{figure} 
    \centering
        \includegraphics[width=\columnwidth]{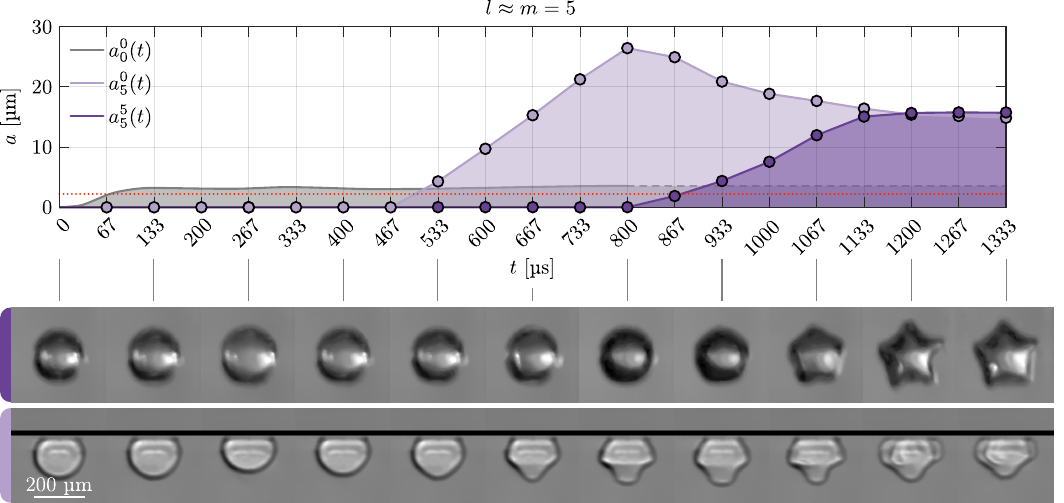}
    \caption{Time evolution of the experimental amplitude of the breathing, zonal and sectoral mode for the case $l \approx m=5$ ($l =4.85$).
    The red dotted line marks the threshold amplitude of the breathing mode for the onset of shape modes.
    The amplitude of the breathing mode is shown as dashed during the occurrence of the sectoral mode, indicating that accurate measurement is not possible due to the loss of bubble axisymmetry.
    Below is a sequence of top-down optical and side-view X-ray images illustrating the dynamics of the bubble.
    The images have been denoised and their backgrounds have been removed for visual clarity.
    The equilibrium bubble radius is $R_0 = \SI{103} {\micro\meter}$.
    The ultrasound driving frequency is $f_{\rm d} = \SI{30}{\kilo\hertz}$ and its amplitude is $p_{\rm a} = \SI{4.4} {\kilo\pascal}$.}
    \label{fig:AnalysisN5}
\end{figure}

\begin{figure} 
    \centering
        \includegraphics[width=\columnwidth]{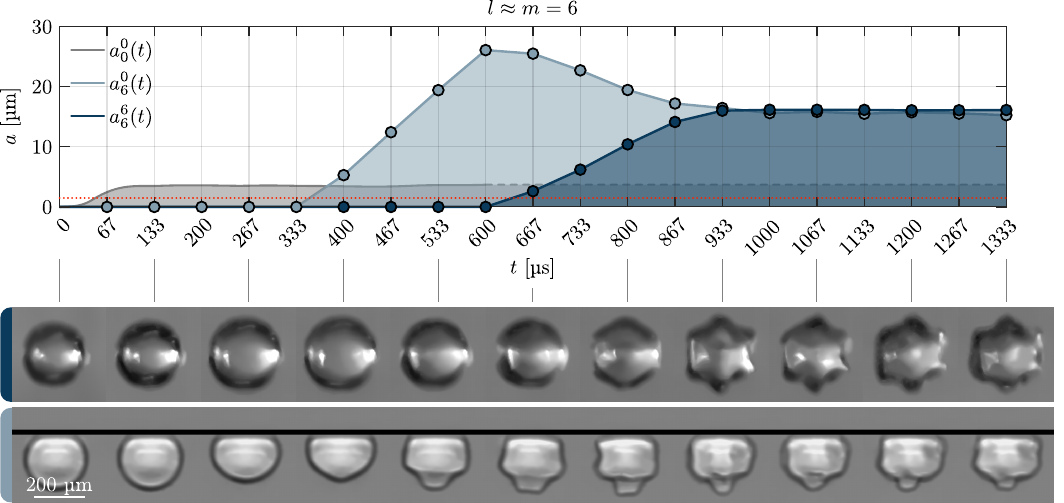}
    \caption{Time evolution of the experimental amplitude of the breathing, zonal and sectoral mode for the case $l \approx m=6$ ($l =6.02$).
    The red dotted line marks the threshold amplitude of the breathing mode for the onset of shape modes.
    The amplitude of the breathing mode is shown as dashed during the occurrence of the sectoral mode, indicating that accurate measurement is not possible due to the loss of bubble axisymmetry.
    Below is a sequence of top-down optical and side-view X-ray images illustrating the dynamics of the bubble.
    The images have been denoised and their backgrounds have been removed for visual clarity.
    The equilibrium bubble radius is $R_0 = \SI{129} {\micro\meter}$.
    The ultrasound driving frequency is $f_{\rm d} = \SI{30}{\kilo\hertz}$ and its amplitude is $p_{\rm a} = \SI{10.5} {\kilo\pascal}$.}
    \label{fig:AnalysisN6}
\end{figure}

The time evolution of the amplitude of bubble deformation components, specifically, the breathing, zonal, and sectoral modes, is presented in figures \ref{fig:AnalysisN3}, \ref{fig:AnalysisN4}, \ref{fig:AnalysisN5}, and \ref{fig:AnalysisN6} for cases with $l \approx m = 3$, $l \approx m = 4$, $l \approx m = 5$, and $l \approx m = 6$, respectively.
The amplitude of the breathing mode $a_0^0(t)$ is approximately measured using the method outlined in section \ref{BrMode}.
This approach remains accurate as long as the bubble maintains an approximately axisymmetric shape.
The amplitude $a_l^0(t)$ of a zonal shape mode $Y_{l}^{0}(\theta,\phi)=P_{l}^{0}(\cos (\theta))$ can be determined by calculating the inner product between the cross section of the bubble deformation, $\eta(\theta,0,t)$, along the $z$-axis, and the shape mode itself:
\begin{equation}\label{eq:InnerProduct4Amplitude}
a_l^0(t) =  \frac{2l+1}{2} \left\langle  \eta(\theta,0,t), P_{l}^{0}(\cos (\theta))\right\rangle = \frac{2l+1}{2}\ \int_{0}^{\alpha_0} \eta(\theta,0,t)  P_{l}^{0} (\cos(\theta)) \sin(\theta) d\theta,
\end{equation}
where $ \frac{2l+1}{2}$ is a normalisation factor that follows from Rodrigues' formula.
The amplitude $a_m^m(t)$ of a sectoral shape mode $Y_{m}^{m}(\theta,\phi)= N_{m}^{m} P_{m}^{m} (\cos (\theta)) \cos(m \phi)$ can be computed in a similar way by computing the inner product between the cross sections along the $x$-axis and $y$-axis of the bubble deformation $\eta(\pi/2,\phi,t)$ and of the shape mode $Y_{m}^{m}(\pi/2,\phi)=\cos(m\phi)$, respectively, as follows:
\begin{equation}
a_m^m(t) = \frac{1}{\pi} \left\langle  \eta(\pi/2,\phi,t), \cos(m\phi)\right\rangle = \frac{1}{\pi}\ \int_{0}^{2\pi} \eta(\pi/2,\phi,t)  \cos(m\phi) d\phi.
\end{equation}
However, under the present conditions, quantifying the amplitude of the zonal mode using equation (\ref{eq:InnerProduct4Amplitude}) is unfeasible.
The non-quantised nature of the shape mode precludes precise measurement of its degree.
Moreover, the bubble deformation, constrained within the range $\theta = \pm \alpha_0$, renders the calculation inherently ill-posed.
Finally, the concurrent presence of a sectoral shape mode complicates the isolation of its specific contribution to the cross-sectional deformation.
To address these challenges, we estimate the amplitude of the shape mode by measuring the amplitude of the lower lobe oscillation, at $\theta = 0$.
By definition, it directly corresponds to the amplitude of the shape mode, and it remains unaffected by sectoral modes, as these modes exhibit zero deformation at \(\theta = 0\).
These issues do not arise in estimating the sectoral mode amplitude due to the quantised nature of the shape mode order, the periodicity of the deformation, and the fact that the zonal mode merely introduces a uniform offset to the deformation in the examined cross section.

In all examined cases, the amplitude of the breathing mode ranges between 1.5 and 2.4 times the threshold required to trigger shape modes for the specific bubble size under investigation.
The temporal evolution of shape modes exhibits consistent patterns across all cases. The zonal mode invariably emerges first, followed by the sectoral mode.
At the breathing mode amplitudes relative to the shape mode threshold employed, the zonal mode typically occurs after 10–14 ultrasound cycles.
A notable exception is observed in the case of $l \approx m = 3$, where the zonal mode initiates after just 4 cycles.
Following the onset of the zonal mode, the sectoral mode typically manifests after 6--10 ultrasound cycles.
The amplitude of the zonal mode increases approximately linearly with time until the inception of the sectoral mode.
At its peak, the zonal mode is 6 to 8 times more intense than the breathing mode.
This factor appears to increase as the interval between the onset of the zonal mode and the initiation of the sectoral mode prolongs.
The pronounced amplification of bubble deformation produced by shape modes arises from their intrinsic capacity to concentrate kinetic energy at specific regions on the bubble surface, the shape mode lobes.
This contrasts with the breathing mode, where kinetic energy and deformation are distributed uniformly across the entire bubble surface.
The concentration of energy at specific points produced by shape modes is the key factor that enables jet formation at low-amplitude ultrasound pressures, as shown in this study.
Upon the onset of the sectoral mode, the amplitude of the zonal mode ceases to increase and begins to decline, stabilising at approximately 0.5-0.7 times its prior peak value.
Simultaneously, the intensity of the sectoral mode grows to a level comparable to that of the zonal mode, indicating a possible energy transfer from the zonal to the sectoral mode.
This decrease in the overall shape mode intensity suggests that the most favourable conditions for jet formation occur when only the zonal mode is active, as this configuration results in the highest deformation amplitude.
For the remainder of the recording period, the zonal and sectoral modes exhibit minimal variation, maintaining a largely stable pattern.
We do not observe significant changes in the amplitude of the breathing mode when the zonal mode develops.
Once the sectoral mode emerges, extracting volume becomes troublesome, as the bubble shape loses axisymmetry.
Therefore, we can only assume that the amplitude remains constant during this regime as well.

The delayed emergence of the ultimate shape mode pattern observed in wall-bounded bubbles contrasts with the behaviour of free bubbles, where the final Faraday pattern appears directly, without transitioning through an intermediate stage of purely axisymmetric shape modes \citep{Cattaneo2025CyclicDelivery}.
One possible explanation is the friction at the contact line, which may restrict the bubble oscillatory motion along the azimuthal direction, thus hindering the onset of surface destabilisation in that direction.
In contrast, motion along the polar direction is not similarly constrained, making zonal modes easier to develop initially.
These zonal modes, superimposed on the breathing-mode spherical oscillation, may then generate sufficient surface deformation to eventually trigger the onset of sectoral modes.
Interestingly, this progression is not unprecedented—it has been observed in sessile droplets on vibrating surfaces, where zonal modes appear first.
Subsequently, the radial oscillation of these axisymmetric waves induces azimuthal waves at the contact line, leading to the formation of sectoral patterns \citep{Vukasinovic2007DynamicsVibration,Panda2023AxisymmetricDrop}.

\subsection{Instability threshold for repeated jets}

When the amplitude of the shape mode exceeds a critical threshold, the inward folding of the bottom shape mode lobe generates a high-velocity jet directed towards the substrate.
After onset, jetting occurs at each cycle of the shape mode.
Given that the shape mode exhibits a half-harmonic behaviour, the jet frequency is half that of the applied ultrasound driving.
Shape mode-induced cyclic jets on bubbles have likely been observed in previous studies that documented repeated jets resulting from surface deformations at driving frequencies spanning hertz \citep{Crum1979SurfaceBubbles}, kilohertz \citep{Prabowo2011SurfaceBubbles}, and megahertz \citep{Vos2011} ranges.

\begin{figure} 
    \centering
        \includegraphics[width=\columnwidth]{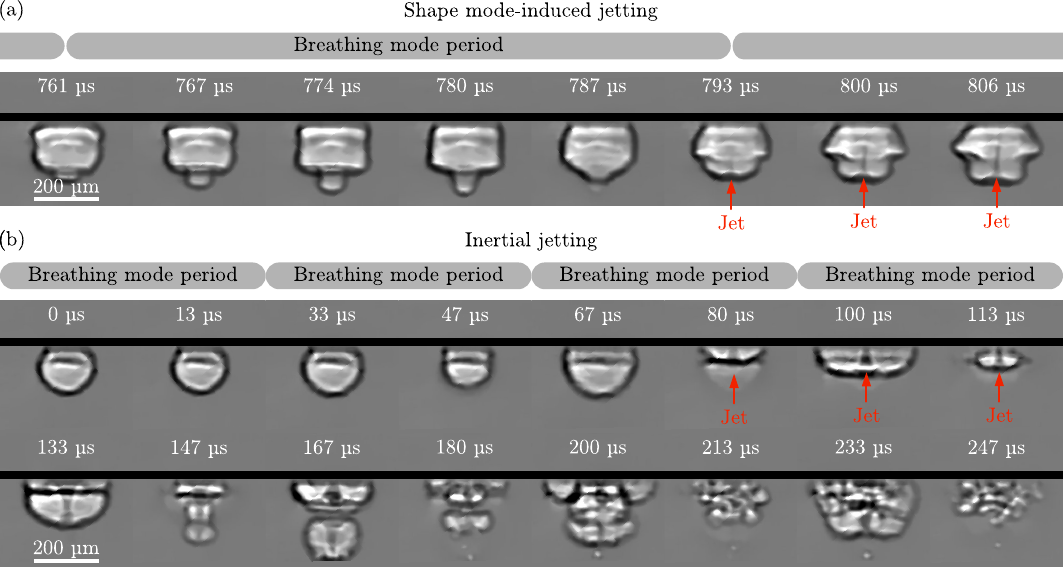}
    \caption{(a) X-ray image sequence of shape mode-induced jetting. 
    The bottom lobe undergoes maximum excursion, and when it rapidly folds inwards, a singularity forms at the fluid interface, where both surface curvature and velocity diverge, leading to the formation of a high-speed jet directed towards the substrate, driven by the self-focusing of kinetic energy at the surface singularity.
    The equilibrium radius of the bubble is $\SI{121}{\micro\meter}$, the driving pressure is $p_{\rm a} = \SI{9.4}{\kilo\pascal}$, and the frequency is $f_{\rm d} = \SI{30}{\kilo\hertz}$.
    (b) X-ray image sequence of ultrasound-driven inertial jetting. 
    The bubble forms a jet as soon as the ultrasound amplitude reaches its steady-state, leaving no time for shape modes to develop.
    This produces a single, quick jet followed by the bubble rapid fragmentation.
    The equilibrium radius of the bubble is $\SI{91}{\micro\meter}$, the driving pressure is $p_{\rm a} = \SI{17}{\kilo\pascal}$, and the frequency is $f_{\rm d} = \SI{30}{\kilo\hertz}$.
    The background of the images has been removed for visual clarity.
    }
    \label{fig:ShapeModeJetting_vs_InertialJetting}
\end{figure}

Figure \ref{fig:ShapeModeJetting_vs_InertialJetting}(a) illustrates that, for a bubble with an equilibrium radius $R_0 = \SI{121}{\micro\meter}$, subjected to a driving pressure $p_{\rm a} = \SI{9.4}{\kilo\pascal}$ and frequency $f_{\rm d} = \SI{30}{\kilo\hertz}$, just before jet emission occurs, the bottom shape mode lobe folds inward, deforming into an approximate truncated cone.
This creates a singularity point at the fluid interface, characterised by divergent surface curvature and velocity.
The resulting self-focusing of kinetic energy at this singularity point initiates the jet formation.
Jets originating from the bottom shape mode lobe are predominantly observed due to its higher amplitude compared to other lobes, as it is not constrained by the substrate.
This contrasts with free bubbles, where all lobes exhibit equal amplitude, allowing simultaneous jet formation when their amplitude is sufficiently high \citep{Cattaneo2025CyclicDelivery}.
This jetting phenomenon can be considered the spherical analog of the jets formed by Faraday waves on vertically-vibrating liquid baths \citep{Longuet-Higgins1983BubblesSurface, Zeff2000SingularitySurface}.
Jets arising from a collapsing conical interface are also observed in other phenomena, including bubble bursting at fluid interfaces \citep{Kientzler1954PhotographicSurface, Duchemin2002JetSurface}, droplet impact on liquid pools \citep{Worthington1897V.Photography, Thoroddsen2018SingularCraters}, cavitation bubble collapse in extremely close proximity to solid boundaries \citep{Lechner2019FastStudy, Reuter2021SupersonicBubbles}, cavitation-bubble-pair interactions \citep{Mishra2022FlowBubbles, Fan2024AmplificationPair}, and bubble coalescence \citep{Jiang2024AbyssCollisions}.

Shape-mode-induced jets are fundamentally different from classical inertial jets, which require considerably higher acoustic pressures to form—almost an order of magnitude more.
Figure \ref{fig:ShapeModeJetting_vs_InertialJetting}(b)  illustrates an example of ultrasound-driven inertial jetting for a bubble with an equilibrium radius $R_0 = \SI{91}{\micro\meter}$, which is near its resonant size.
The bubble is subjected to a driving pressure $p_{\rm a} = \SI{17}{\kilo\pascal}$ and frequency $f_{\rm d} = \SI{30}{\kilo\hertz}$.
Here, the bubble jets immediately upon reaching the steady-state ultrasound amplitude, leaving no time for shape modes to develop.
The result is a single, swift jet followed by rapid fragmentation of the bubble.
Unlike jets formed by shape modes, this jet arises from the intense pressure gradient applied to the bubble rather than from an instability at the interface.
Thus, shape-mode-induced jets represent a unique mechanism for concentrating energy through relatively low acoustic excitation, marking a distinct contrast to the behaviour of classical inertial jets.
\begin{figure} 
    \centering
        \includegraphics[width=\columnwidth]{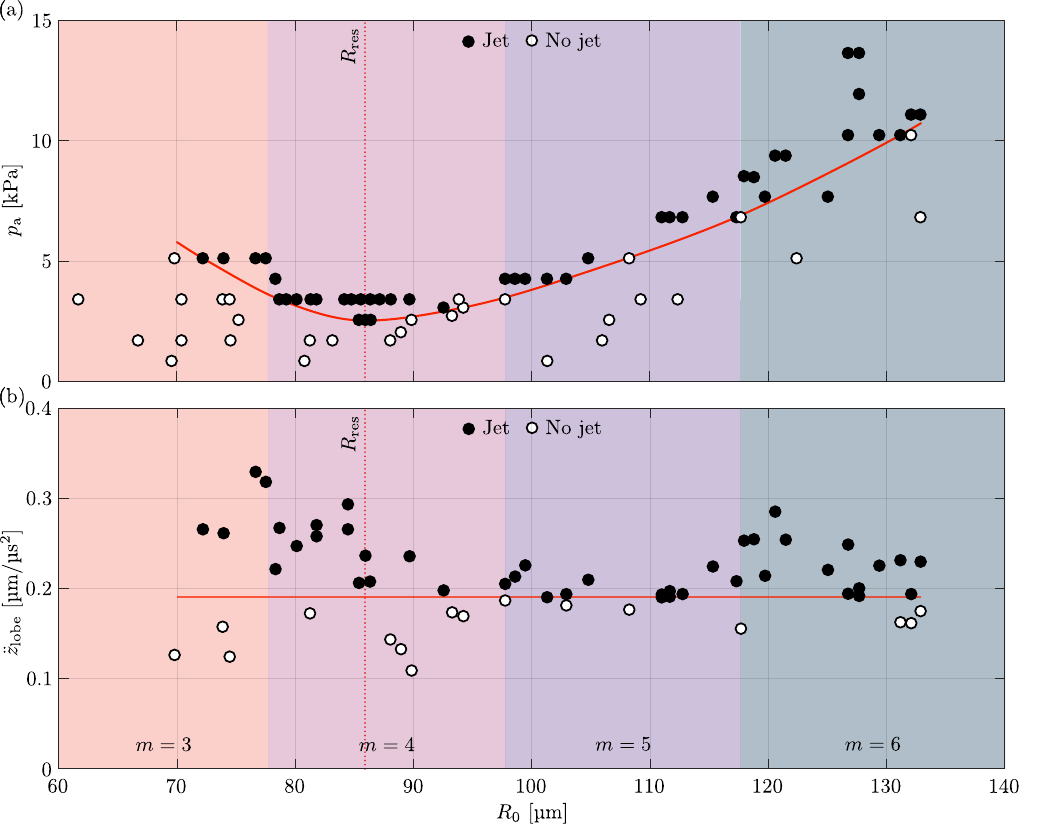}
    \caption{(a) Jetting occurrence as function of bubble size, shape mode order, and driving ultrasound pressure.
    A minimum in driving pressure occurs at the resonant radius.
    (b) Jetting occurrence as function of bubble size, shape mode order, and acceleration of the bottom shape mode lobe.
    A distinct threshold in acceleration, regardless of bubble size, is identified. 
    }
    \label{fig:JettingThresholdAdimensional}
\end{figure}

In figure \ref{fig:JettingThresholdAdimensional}(a), we illustrate the relationship between bubble size and the ultrasound pressure needed to cause shape mode-induced jet formation.
The required pressure reaches a minimum at the resonant radius for the applied driving frequency.
In figure \ref{fig:JettingThresholdAdimensional}(b), we identify a distinct threshold in the acceleration of the bottom shape mode lobe of approximately $\SI{0.2}{\micro\meter\per\micro\second\squared}$, irrespective of the bubble size, beyond which jetting occurs.
The lobe acceleration $\ddot z_{\rm lobe}$ is estimated from the zonal shape mode amplitude as $\ddot z_{\rm lobe} \approx \omega_{0,l}^2 a_{l}^0$, where $\omega_{0,l} = {\omega_{\rm {d}}}/{2}$ is the angular frequency of the shape mode.
This consistent acceleration threshold arises because Faraday instability-driven jetting depends solely on the motion of individual shape mode lobes, which is independent of bubble size.
While increasing the bubble size increases the number of lobes, their individual width—set by the driving frequency—remains unchanged.
As a result, the critical lobe height and corresponding acceleration required for jetting remain constant across bubble sizes.
This behaviour is fundamentally distinct from that of inertial jets, where jet strength scales with the pressure difference across the bubble, and therefore depends both on the magnitude of the pressure gradient and the bubble size \citep{Supponen2016ScalingBubbles}.

The speed of jets generated by shape modes is found to reach $u_{\rm jet} = \SI{30}{\meter\per\second}$.
Using the fluid hammer pressure formula, $p_{\rm wh} = \rho_{\rm l} c_{\rm l} u_{\rm jet}$, for a water flow impinging on a rigid material, the estimated impact pressure on the substrate is calculated to be up to \SI{45}{\mega\pascal}.
This suggests that even at relatively low ultrasound driving pressures, the repeated jet formations are sufficiently powerful to cause damage to biological and ductile materials.
This mechanism may hold significant importance in ultrasound-based cleaning applications and, as demonstrated in our previous work, plays a pivotal role in sonoporation for targeted drug delivery \citep{Cattaneo2025CyclicDelivery}.
As it displaces, the jet may eventually break into multiple droplets following the Rayleigh–Plateau instability.
Jetting persists indefinitely unless the emergence of the sectoral shape mode dampens the lobe amplitude below the critical threshold, or the bubble shape becomes too chaotic as a result of the disruptive influence of jets on the shape itself.

\section{Numerical simulations}\label{Sec6}
We now complement our experimental results with three-dimensional boundary element method (BEM) numerical simulations \citep[e.g.][]{Wang:2014a}, modelling the evolution of the bubble interface and formation of jets.
Full details of the BEM model employed are provided in Appendix \ref{appF}.
\begin{figure} 
    \centering
        \includegraphics[width=\columnwidth]{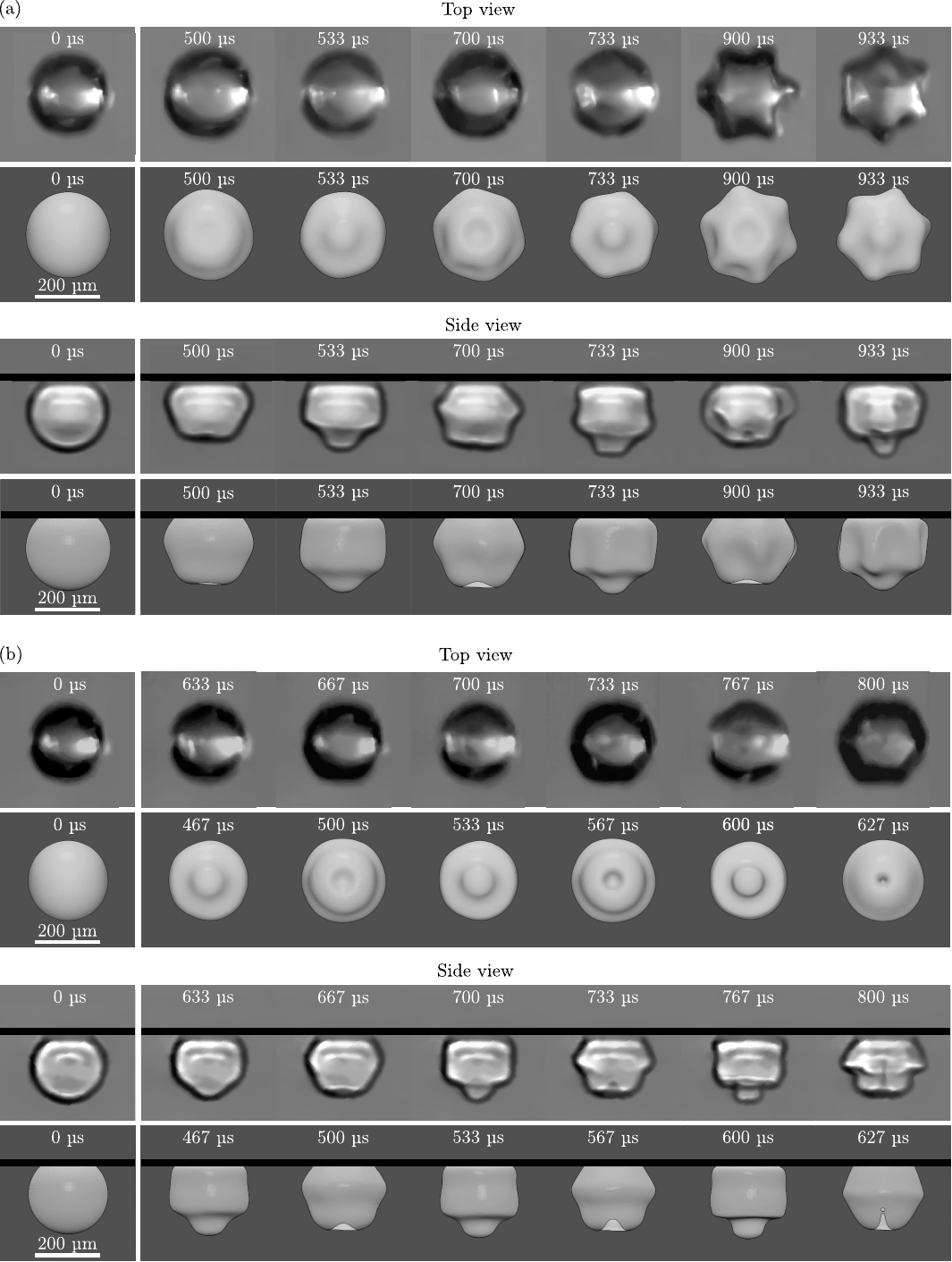}
    \caption{Comparison of the time evolution of the bubble interface between experimental observations and numerical simulations using the boundary element method, shown from both top and side view perspectives. 
    (a) Transition from a purely zonal shape mode to a combined zonal and sectoral mode.
    The bubble has an equilibrium radius $R_0 = \SI{120}{\micro\meter}$ and is driven by ultrasound with an amplitude $p_{\rm a} = \SI{9.4}{\kilo\pascal}$ at a frequency $f_{\rm d} = \SI{30}{\kilo\hertz}$.
    (b) Jetting formation during an almost purely zonal shape mode.
    The bubble has an equilibrium radius $R_0 = \SI{129}{\micro\meter}$ and is driven by ultrasound with an amplitude $p_{\rm a} = \SI{10.5}{\kilo\pascal}$ at a frequency $f_{\rm d} = \SI{30}{\kilo\hertz}$.
    }
    \label{fig:EXPvBIM}
\end{figure}
\begin{figure} 
    \centering
        \includegraphics[width=\columnwidth]{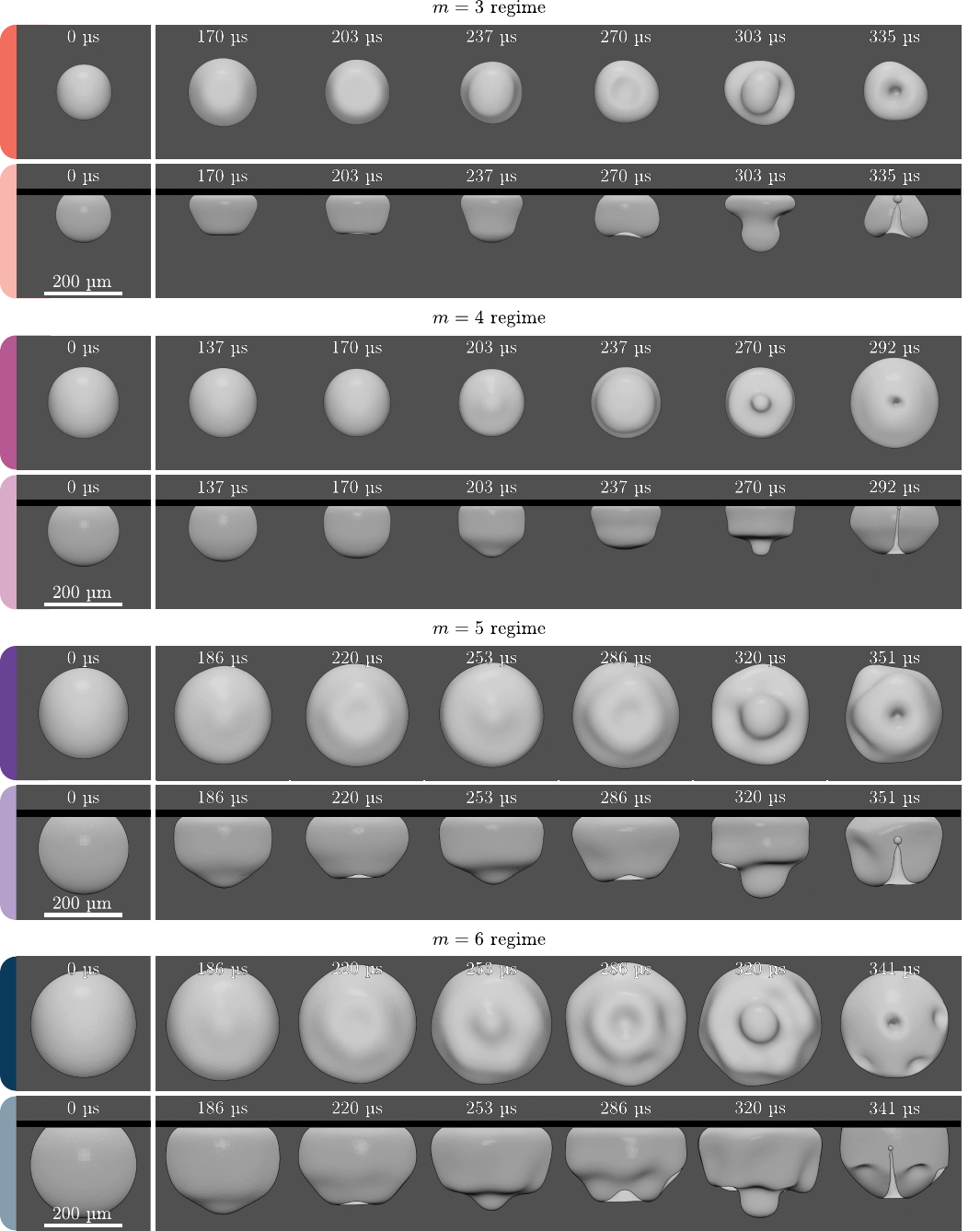}
    \caption{Simulated examples of shape mode-induced jetting, modelled using the boundary element method, presented in both top and side views for various bubble sizes corresponding to the following regimes: (a) $m = 3$, (b) $m = 4$, (c) $m = 5$, and (d) $m = 6$.
    Images are shown at intervals of one per ultrasound cycle, resulting in an interframe time of \SI{33.3} {\micro\second}, except for the last image, which captures the last moment before numerical mesh rupture occurs following jet formation.
    The bubbles have equilibrium radius $R_0$ of \SI{70} {\micro\meter}, \SI{90} {\micro\meter}, \SI{115} {\micro\meter}, and \SI{135} {\micro\meter}, and are subjected to pressure amplitudes $p_{\rm a}$ of \SI{18} {\kilo\pascal}, \SI{3.6} {\kilo\pascal}, \SI{14} {\kilo\pascal}, and \SI{24} {\kilo\pascal}, respectively, at a frequency $f_{\rm d}$ of \SI{30} {\kilo\hertz}.
    }
    \label{fig:BIM_Simulations}
\end{figure}
Figure \ref{fig:EXPvBIM} presents a sequence of phase-matched frames comparing the experimentally observed and BEM-calculated time evolution of the bubble interface.
In the first subfigure, we examine the transition from a purely zonal mode to a combined zonal and sectorial mode for a bubble with an equilibrium radius of $R_0 = \SI{129}{\micro\meter}$, subjected to ultrasound driving at an amplitude $p_{\rm a} = \SI{10.5}{\kilo\pascal}$ and a frequency $f_{\rm d} = \SI{30}{\kilo\hertz}$.
The simulation closely matches the observed data, both in terms of the onset of the transition, occurring at around $t=\SI{500}{\micro\second}$ (15 ultrasound cycles), and the transition duration, which spans about $\SI{400}{\micro\second}$.
Moreover, throughout the entire transition, the amplitude of the various shape modes exhibits good agreement between the simulation and experimental results.
In the second subfigure, we examine the formation of a jet during an almost purely zonal shape mode—the most common manifestation, as previously observed—for a bubble with an equilibrium radius of $R_0 = \SI{120}{\micro\meter}$, subjected to ultrasound driving at an amplitude $p_{\rm a} = \SI{9.4}{\kilo\pascal}$ and a frequency $f_{\rm d} = \SI{30}{\kilo\hertz}$.
The simulation terminates when a drop detaches from the first ejected jet due to the Rayleigh-Plateau instability, causing the mesh to break up.
Up to that point, the simulation and experimental observations of the bubble interface evolution are in good agreement.
Note that in the final time instant compared, the discrepancy between the experimental and simulation results, especially clear in the jet displacement, arises because the two frames are not phase-matched as the simulation concludes \SI{6}{\micro\second} earlier.
Although the simulation predicts the emission of the first jet five ultrasound cycles earlier than observed experimentally, this discrepancy is acceptable given the inherent variability in the number of cycles required to produce a jet in the experiments.
In conclusion, these results support the use of the BEM for simulating the dynamics of wall-attached ultrasound-driven bubbles of this scale.

Figure \ref{fig:BIM_Simulations} displays examples of simulated shape mode-induced jetting, modelled using the BEM, for various bubble sizes corresponding to the $m = 3$, $m = 4$, $m = 5$, and $m = 6$ regimes.
The simulation stops when the jet either touches the substrate or pinches off a droplet, disrupting the mesh and thereby restricting the simulation to display only the first jet occurrence.
We select ultrasound pressures to induce shape mode-induced jet formation within a similar number of ultrasound cycles across different cases.
For the representative bubble in the $m=4$ regime, a pressure of just $p_{\rm a} = \SI{3.6}{\kilo\pascal}$ over 8 ultrasound cycles is sufficient to initiate jetting.
This is because the bubble radius, $R_0 =\SI{90}{\micro\meter}$, nearly matches the resonant radius. 
In the other regimes, higher ultrasound pressures are needed to achieve jet formation within a similar cycles count: $p_{\rm a} =\SI{14}{\kilo\pascal}$ for the $m=5$ case ($R_0 =\SI{115}{\micro\meter}$), $p_{\rm a} =\SI{18}{\kilo\pascal}$ for the $m=3$ case ($R_0 =\SI{70}{\micro\meter}$), and $p_{\rm a} =\SI{24}{\kilo\pascal}$ for the $m=6$ case ($R_0 =\SI{135}{\micro\meter}$).
For these bubble sizes, jetting can occur even at lower pressures, but initiating it requires more cycles.
In any case, the pressure needed to generate jets through shape modes is significantly lower than the pressure required for classical inertial jets.

\section{Conclusions}

In this study, we experimentally investigated the behaviour of individual micrometric air bubbles resting against a rigid substrate when subjected to ultrasound driving, focusing specifically on the development of interfacial instabilities over time.
To achieve this, we developed a dual-view imaging setup that integrates a visible light top view with a phase-contrast X-ray side view, powered by the synchrotron radiation source at the European Synchrotron Research Facility (ESRF) on beamline ID19.
This system enables high-speed, high-magnification simultaneous recordings from both perspectives, providing unprecedented insights into the dynamic evolution of bubble shapes under ultrasound driving.

The observed bubble dynamics can be organised into four distinct consecutive regimes:
(i) the initial purely spherical oscillation, also referred to as breathing mode.
(ii) the onset of harmonic axisymmetric meniscus waves,
(iii) the emergence of a higher-amplitude half-harmonic axisymmetric mode, also referred to as zonal mode, triggered by the Faraday instability, and
(iv) the superposition of a half-harmonic sectoral mode onto the zonal mode.
Notably, the stepwise transition through distinct shape modes regimes contrasts with the behaviour of free bubbles, which exhibit their ultimate Faraday wave pattern immediately upon the onset of instability.

When a lobe of the Faraday-induced shape mode folds inward with sufficient intensity during its oscillatory motion, it generates a jet directed towards the substrate.
This jet formation occurs at mild ultrasound pressures and recurs with each cycle of the shape mode.
These jets, arising from interface instabilities, represent a new and distinct class of bubble-generated jets. 
They fundamentally differ from classical inertial jets, which form under high-pressure gradients during the bubble collapse phase and require significantly higher acoustic pressures.

Our detailed analysis of each regime reveals several important insights.
To accurately model the bubble volume evolution over time, the commonly used polytropic gas approximation in bubble dynamics is insufficient.
Accurate predictions are only achieved with a more advanced model that accounts for thermal damping effects.
The observed harmonic axisymmetric pattern is driven by meniscus waves originating from the contact line.
These waves are equivalent to those found on flat water-air interfaces in containers with a wall contact angle different from $90^\circ$, when subjected to vertical oscillations.
The shape mode spectrum is derived through a kinematic analysis under various wetting conditions.
For pinned contact line or fixed contact angle boundary conditions, the spectrum is quantised and non-degenerate, whereas for free boundary conditions, the spectrum becomes degenerate and exhibits a continuous range of shape mode degrees.
Our experimental results, consistent with the non-restricting boundary conditions imposed by the chosen substrate, confirm this expected continuity in shape mode degrees and the observed degeneracy of the shape modes.
This contrasts with free bubbles, which, while also exhibiting a degenerate spectrum, display only discrete shape mode degrees constrained by their spherical periodicity.

Furthermore, we find a strong correlation between the experimentally measured ultrasound pressure thresholds for the onset of shape mode and the theoretical predictions based on the model from \cite{Francescutto1978PulsationBubbles}, when using a lowered surface tension value of \SI{65}{\milli\newton\per\meter}.
This reduction in surface tension might be attributed to the presence of surfactants on the bubble surface.
Our analysis of the time evolution of breathing, zonal, and sectoral mode amplitudes reveals that shape modes reach significantly higher amplitudes compared to the breathing mode.
This suggests that shape modes are more effective at amplifying acoustic energy and concentrating it in localised regions, which explains the formation of shape mode-driven jets under moderate acoustic pressures.
We also identify a critical acceleration threshold for shape mode lobes, independent of bubble size, beyond which jetting occurs.
Unlike in free bubbles, where jets may form from any point on the surface, jetting in wall-attached bubbles consistently emerges from the side not constrained by the substrate.

Finally, we supplement our experimental results with three-dimensional simulations of bubble dynamics using the boundary element method which is based on the assumption of potential flow.
These simulations align closely with the experimental visualisations, accurately capturing the evolution of bubble shape, the events timing and their required ultrasound pressure, in both the transition from zonal to mixed modes (zonal and sectoral) and the formation of jets following the emergence of strong zonal modes.
Our results support the use of boundary element method as a reliable approach for modelling the dynamics of ultrasound-driven wall-attached bubbles at this scale.

In conclusion, our study provides valuable insights into the dynamics of wall-attached bubbles under ultrasound driving, advancing our understanding of the development of surface instabilities and jet formation.
This research has significant implications for industrial and biomedical applications, such as ultrasound cleaning and bubble-assisted targeted drug delivery. \\

\noindent\textbf{Supplementary data} {\label{SupMat}The supplementary movie is available at \url{https://doi.org/10.1017/jfm.2025.10457} \\

\noindent\textbf{Acknowledgements} {The results presented in this work were obtained during the allocated beamtime for proposal ME-1633 on beamline ID19 at the European Synchrotron Radiation Facility (ESRF).} \\

\noindent\textbf{Funding} {This work is supported by ETH Z\"urich and the European Synchrotron Radiation Facility (ESRF).} \\

\noindent\textbf{Declaration of interests} {The authors report no conflict of interest.} \\

\noindent\textbf{Data availability statement} {The data that support the findings of this study are available under reasonable request.} \\

\noindent\textbf{Author ORCIDs}{
\\ \orcidlink{https://orcid.org/0000-0002-1581-4280} Marco Cattaneo \href{https://orcid.org/0000-0002-1581-4280}{https://orcid.org/0000-0002-1581-4280} 
\\ \orcidlink{https://orcid.org/0009-0005-1932-5846} Louan Presse \href{https://orcid.org/0009-0005-1932-5846}{https://orcid.org/0009-0005-1932-5846} 
\\ \orcidlink{https://orcid.org/0000-0001-7747-2012} Gazendra Shakya \href{https://orcid.org/0000-0001-7747-2012}{https://orcid.org/0000-0001-7747-2012}
\\ \orcidlink{https://orcid.org/0009-0007-1064-7544} Thomas Renggli
\href{https://orcid.org/0009-0007-1064-7544}{https://orcid.org/0009-0007-1064-7544}
\\ \orcidlink{https://orcid.org/0000-0001-9069-9246} Bratislav Lukić \href{https://orcid.org/0000-0001-9069-9246}{https://orcid.org/0000-0001-9069-9246}
\\ \orcidlink{https://orcid.org/0009-0004-5687-8936} Anunay Prasanna \href{https://orcid.org/0009-0004-5687-8936}{https://orcid.org/0009-0004-5687-8936}
\\ \orcidlink{https://orcid.org/0000-0002-8311-2985} Daniel W. Meyer
\href{https://orcid.org/0000-0002-8311-2985}{https://orcid.org/0000-0002-8311-2985}
\\ \orcidlink{https://orcid.org/0000-0001-9486-3621} Alexander Rack \href{https://orcid.org/0000-0001-9486-3621}{https://orcid.org/0000-0001-9486-3621}
\\ \orcidlink{https://orcid.org/0000-0001-6738-0675} Outi Supponen \href{https://orcid.org/0000-0001-6738-0675}{https://orcid.org/0000-0001-6738-0675}
}\\

\noindent\textbf{Author contributions} {M.C., G.S., and O.S. conceived the study and secured beam time for the X-ray experiments.
M.C., L.P., G.S., B.L., and O.S. defined the experimental methodology.
M.C., L.P., G.S., B.L., A.P., and O.S. performed the experiments.
M.C. and L.P. post-processed and analysed the data.
M.C. carried out the theoretical modelling.
T.R. and D.W.M. performed the BEM simulations.
M.C. wrote the initial draft of the paper.
O.S., B.L., and A.R. provided resources.
O.S. supervised the research.
All authors contributed to the critical review and revision of the manuscript.}\\

\appendix

\section{Derivation of the variational wetting relation}\label{appA}
A perturbation of the bubble surface $\delta{\boldsymbol{R}}$ induces a change in the wetting condition:
\begin{equation}\label{eq:YD_pert}
\delta ({\boldsymbol{n}} \cdot {\boldsymbol{n}}_{\rm s})|_{\theta = \alpha_0} = (\delta_{\perp} {\boldsymbol{n}} + \delta_{\parallel} {\boldsymbol{n}})|_{\theta = \alpha_0} \cdot {\boldsymbol{n}}_{\rm s} = -\sin(\alpha_0) \delta \alpha,
\end{equation}
where $\delta_{\perp} {\boldsymbol{n}}$ and $\delta_{\parallel} {\boldsymbol{n}}$ represent the variations due to the normal $\eta$ and tangential $\tau$ components of the surface deformation, respectively. 
The term $\delta\alpha$ denotes the change in the contact angle relative to the equilibrium state.
Since the substrate is flat, $\delta {\boldsymbol{n}}_{\rm s} = 0$.
The normal unit vector is given by:
\begin{equation}
{\boldsymbol{n}} = \frac{\partial_{\theta} {\boldsymbol{R}} \times \partial_{\phi} {\boldsymbol{R}}}{\| \partial_{\theta} {\boldsymbol{R}} \times \partial_{\phi} {\boldsymbol{R}} \|},
\end{equation}
where $ \partial_{\theta} {\boldsymbol{R}} = \frac{\partial {\boldsymbol{R}}}{\partial \theta}$ and $ \partial_{\phi} {\boldsymbol{R}} = \frac{\partial {\boldsymbol{R}}}{\partial \phi}$.
Considering the normal component of the surface deformation $\delta_{\perp} {\boldsymbol{R}} = \eta {\boldsymbol{n}}$, we obtain for the variation $\delta_{\perp} {\boldsymbol{n}}$:
\begin{multline}
\delta_{\perp} {\boldsymbol{n}} = \frac{1}{\| \partial_{\theta} {\boldsymbol{R}} \times \partial_{\phi} {\boldsymbol{R}} \|}\left(\partial_{\theta}\eta{\boldsymbol{n}} \times  \partial_{\phi} {\boldsymbol{R}} +  \partial_{\theta} {\boldsymbol{R}}  \times \partial_{\phi}\eta{\boldsymbol{n}} \right) + \\
+ \frac{\eta}{\| \partial_{\theta} {\boldsymbol{R}} \times \partial_{\phi} {\boldsymbol{R}} \|} \left(   \partial_{\theta}{\boldsymbol{n}} \times  \partial_{\phi} {\boldsymbol{R}}  +  \partial_{\theta} {\boldsymbol{R}}  \times \partial_{\phi} {\boldsymbol{n}}  \right) - \frac{\delta \| \partial_{\theta} {\boldsymbol{R}} \times \partial_{\phi} {\boldsymbol{R}} \|}{\| \partial_{\theta} {\boldsymbol{R}} \times \partial_{\phi} {\boldsymbol{R}} \|}{\boldsymbol{n}}.
\end{multline}
The first term on the right-hand side is a vector orthogonal to ${\boldsymbol{n}}$, while the sum of the remaining two terms constitutes a vector parallel to ${\boldsymbol{n}}$.
Given that the variation of a unit vector is orthogonal to the vector itself, the sum of the last two terms must be zero.
Furthermore, since the angular coordinates $\theta$ and $\phi$ correspond numerically to the lengths of their respective arcs, it follows that  $\| \partial_{\theta} {\boldsymbol{R}} \times \partial_{\phi} {\boldsymbol{R}} \| = 1$. 
Therefore:
\begin{equation}\label{eq:delta_perp_n}
\delta_{\perp} {\boldsymbol{n}} = -  \partial_{\theta}\eta {\boldsymbol{t}}  -  \partial_{\phi}\eta {\boldsymbol{b}}.
\end{equation}
Considering the tangential component of the surface deformation $\delta_{\parallel} {\boldsymbol{R}} = \tau {\boldsymbol{t}}$, we derive the variation $\delta_{\parallel} {\boldsymbol{n}}$ by leveraging the directional derivative of the bubble surface:
\begin{equation}
\delta_{\parallel} {\boldsymbol{n}} =  \partial_{\theta} {\boldsymbol{n}}\tau.
\end{equation}
Using the Frenet--Serret formulas, $ \partial_{\theta} {\boldsymbol{n}} = - k{\boldsymbol{t}}$, where $k = -\|{\boldsymbol{R}}_0\|^{-1} = -1$ is the curvature of the bubble surface along the tangential direction ${\boldsymbol{t}}$, we obtain:
\begin{equation}\label{eq:delta_par_n}
\delta_{\parallel} {\boldsymbol{n}} = \tau {\boldsymbol{t}}.
\end{equation}
By inserting equations (\ref{eq:delta_perp_n}) and (\ref{eq:delta_par_n}) in equation (\ref{eq:YD_pert}) and considering that ${\boldsymbol{t}} \cdot {\boldsymbol{n}}_{\rm s} = - \sin(\alpha_0)$, ${\boldsymbol{b}} \cdot {\boldsymbol{n}}_{\rm s} = 0$ and $\tau|_{\theta=\alpha_0} = \cot(\alpha_0) \eta|_{\theta=\alpha_0}$, we obtain the following variational wetting relation (equation~\ref{eq:vwr}):
\begin{equation}
 \partial_{\theta}\eta|_{\theta=\alpha_0} - \cot(\alpha_0)\eta|_{\theta=\alpha_0} = -\delta\alpha.
\end{equation}

\section{Visualisation of zonal shape modes for different bubble boundary conditions}\label{appB}
\renewcommand{\thefigure}{B\arabic{figure}}
\setcounter{figure}{0}
\begin{figure} 
    \centering 
    \includegraphics[width=\columnwidth]{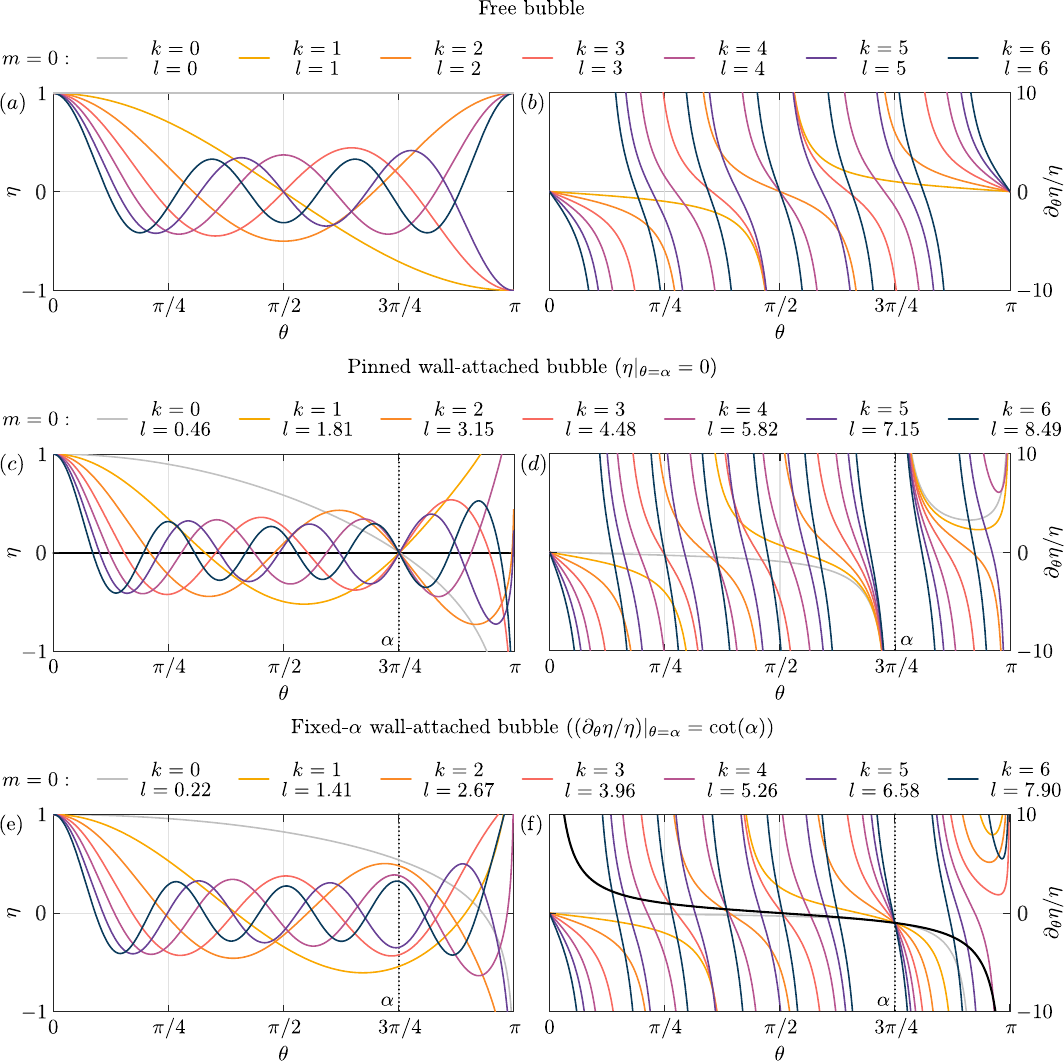}
    \caption{
    First seven shape modes $P_{l_{k}}^m(\cos(\theta))$ for $m=0$ describing the normal bubble deformation $\eta$ for: (a-b) a free bubble, (c-d) a pinned bubble with a static contact angle $\alpha_0 = 3\pi/4$, and (e-f) a fixed-contact-angle bubble with $\alpha_0 = 3\pi/4$.
    }
    \label{fig:FractionalLegendreFunction}
\end{figure}
Figure~\ref{fig:FractionalLegendreFunction} presents the first seven shape modes \( P_{l_k}^m(\cos(\theta)) \) for \( m = 0 \), which describe the normal deformation \( \eta \) of a bubble surface.
The integer \( k \) sequentially orders the permissible modes at a given \( m \).
Panels (a) and (b) show the shape modes for a free bubble.
In this case, to comply with 2$\pi$-periodicity of $\eta$ and its directional derivatives, both \( l \) and \( m \) must be integers.
Consequently, $l$ corresponds with $k$.
Panels (c) and (d) illustrate the shape modes for a pinned bubble with a static contact angle \( \alpha_0 = \frac{3\pi}{4} \).
Here, the surface deformation must vanish at the angle \( \theta = \alpha_0 \).
This boundary condition requires \( l \) to be typically non-integer, meaning it no longer corresponds to \( k \).
The zero-deformation line is represented by a bold black line.
Panels (e) and (f) depict the shape modes for a bubble with a fixed contact angle of \( \alpha_0 = \frac{3\pi}{4} \).
In this scenario, the ratio between the surface deformation and its directional derivative along \( \theta \) at \( \theta = \alpha_0 \) must equal \( \cot(\alpha_0) \).
Like the pinned case, this condition typically results in non-integer values of \( l \), which again do not match \( k \).
The \( \cot(\alpha_0) \) line is shown as a bold black line.

\section{Permissible values of shape mode degrees for different bubble boundary conditions}\label{appG}

The table \ref{tab:kd} lists the permissible values of the degree \( l \) of a shape mode for given indices \( k \) and orders \( m \), in two different cases. 
Case (a) corresponds to a pinned wall-attached bubble with a static contact angle \( \alpha_0 = \frac{3\pi}{4} \), while case (b) corresponds to a fixed-contact-angle wall-attached bubble with \( \alpha_0 = \frac{3\pi}{4} \).
The values shown for case (a) correspond to the shape mode spectrum depicted in figure~\ref{fig:Spectrum}(b), and those for case (b) correspond to the shape mode spectrum depicted in figure~\ref{fig:Spectrum}(c).

\renewcommand{\thetable}{C\arabic{table}}
\setcounter{table}{0}
\begin{table}
  \begin{center}
\setlength\aboverulesep{0ex}
\setlength\belowrulesep{0px}
\def~{\hphantom{0}}
      \begin{tabular}{cccccccc}
      \multicolumn{8}{c}{(a) \qquad    Pinned wall-attached bubble \ \    \qquad \qquad}      \vspace{0.2cm}\\
      \diagbox[width=0.75cm, height=0.75cm]{k}{m}
          &0    &1    &2    &3    &4    &5    &6   \\
      0   &0.46 &     &     &     &     &     &    \\
      1   &1.81 &1.24 &     &     &     &     &    \\
      2   &3.15 &2.55 &2.14 &     &     &     &    \\
      3   &4.48 &3.87 &3.37 &3.08 &     &     &    \\
      4   &5.82 &5.19 &4.66 &4.25 &4.04 &     &    \\
      5   &7.15 &6.52 &5.96 &5.49 &5.17 &5.02 &    \\
      6   &8.49 &7.85 &7.27 &6.76 &6.37 &6.11 &6.01\\
      \end{tabular}
      \quad \quad
      \begin{tabular}{cccccccc}
      \multicolumn{8}{c}{(b) \quad  \ \ \ \  Fixed-$\alpha$ wall-attached bubble \ \ \ \ \ \  \quad \qquad}      \vspace{0.2cm}\\
      \diagbox[width=0.75cm, height=0.75cm]{k}{m}
          &0    &1    &2    &3    &4    &5    &6   \\
      0   &0.22 &     &     &     &     &     &    \\
      1   &1.41 &1.00 &     &     &     &     &    \\
      2   &2.67 &2.14 &1.95 &     &     &     &    \\
      3   &3.96 &3.37 &3.00 &2.96 &     &     &    \\
      4   &5.26 &4.66 &4.18 &3.94 &3.97 &     &    \\
      5   &6.58 &5.96 &5.43 &5.05 &4.92 &4.98 &    \\
      6   &7.90 &7.27 &6.71 &6.25 &5.96 &5.93 &5.99\\
      \end{tabular}
  \caption{Permissible values of the degree $l$ of a shape mode for a given set of indexes $k$ and orders $m$ in the cases of, (a) a pinned wall-attached bubble with a static contact angle $\alpha_0 = 3\pi/4$, and (b) a fixed-contact-angle wall-attached bubble with $\alpha_0 = 3\pi/4$.
  }
  \label{tab:kd}
  \end{center}
\end{table}

\section{Derivation of the pressure term accounting for the rigid boundary using the method of images}\label{appC}

The radial motion of a spherical bubble of radius $R(t)$ in an unbounded  medium under spatially uniform acoustic forcing can be described using the Rayleigh–Plesset equation for mildly compressible Newtonian media, which reads:
\begin{equation}\label{eq:RP}
\rho_{\rm l} \left(R \ddot R + \frac{3}{2} \dot R^2 \right)= \left( 1 + \frac{R}{c_{\rm l}} \frac{d}{dt} \right) p_{\rm g}
   -2\frac{\sigma_{\rm lg}}{R} - p_{\infty} - p_{\rm d}\left(t\right) - 4\mu_{\rm l} \frac{\dot R}{R},
\end{equation}
To incorporate the influence of the solid boundary, we employ the method of images \citep{Strasberg1953TheLiquids}.
This approach, rooted in the potential flow theory, simulates a flat boundary within the flow field by introducing a virtual bubble mirroring the real bubble on the opposite side of the boundary.
Due to the linear nature of the potential flow approximation, the influence of each bubble on the flow field can be analysed independently.
Following the approach adopted in several other studies \citep{Zabolotskaya1984INTERACTIONFIELD.,Oguz1990AOscillations,Mettin1997BjerknesField,Pelekasis2004SecondaryStreamers,Bremond2006ControlledCavitation}, the velocity field $u_{\rm l}(r,t)$ induced by each bubble can be derived from the continuity equation and is expressed as:
\begin{equation}
u_{\rm l}(r,t) = \frac{R^2 \dot{R}}{r^2},
\end{equation}
where $r$ is the radial coordinate, with the origin of the coordinate system at the centre of the bubble.
The corresponding pressure field $p_{\rm l}(r,t)$ can be derived from the momentum equation:
\begin{equation}
\rho_{\rm l}\frac{\partial u_{\rm l}}{\partial t} = - \frac{\partial p_{\rm l}}{\partial r},
\end{equation}
where the nonlinear term is omitted due to its relatively smaller magnitude compared to the other terms.
Integrating this equation yields the pressure field:
\begin{equation}
p_{\rm l}(r,t) = \rho_{\rm l} \frac{2R\dot{R}^2 + R^2 \ddot{R}}{r},
\end{equation}
which acts as an additional driving pressure for the second bubble.
Therefore the equation governing the radial motion of a bubble in proximity to a solid boundary can be approximated as follows (equation \ref{eq:RPplusImage}):
\begin{equation}
\rho_{\rm l} \left(R \ddot R + \frac{3}{2} \dot R^2 \right)= \left( 1 + \frac{R}{c_{\rm l}} \frac{d}{dt} \right) p_{\rm g}
   -2\frac{\sigma_{\rm lg}}{R} - p_{\infty} - p_{\rm d}\left(t\right) - \rho_{\rm l} \frac{2R\dot{R}^2 + R^2 \ddot{R}}{2d} - 4\mu_{\rm l} \frac{\dot R}{R},
\end{equation}

\section{Zhou model for the bubble gas pressure}\label{appD}

The method is based on the assumption that the gas pressure within the bubble is uniform \citep{Prosperetti1988NonlinearDynamics}.
For an ideal gas, the radial velocity $u_{\rm g}(r)$ is expressed as:
\begin{equation}
u_{\rm g}(r) = \frac{1}{\gamma p_{\rm g}}\left(\left(\gamma-1\right) K_{\rm g} \frac{\partial T_{\rm g}}{\partial r} -\frac{1}{3} r \dot{p}_{\rm g} \right),
\end{equation}
leading to an exact relation for the gas pressure:
\begin{equation}
\dot{p}_{\rm g} = \frac{3}{R}\left(\left(\gamma-1\right) K_{\rm g} \left.\frac{\partial T_{\rm g}} {\partial r}\right|_{R} -\gamma p_{\rm g} \dot{R} \right),
\end{equation}
where $r$ is the radial coordinate, $\gamma$ is the gas specific heat ratio, $K_{\rm g}$ is the gas thermal conductivity, and $T_{\rm g}$ is the gas temperature.
The temperature profile $T_{\rm g}(r)$ inside the bubble is divided into three regions: (1) an internal layer with uniform temperature, (2) a buffer layer, and (3) an external layer with a linear temperature distribution.
The change in bubble surface temperature is negligible, thus $T_{\rm g}|_{R} \approx T_{\rm l}$ \citep{Prosperetti1988NonlinearDynamics}. 
The volume-averaged temperature $T_{\rm g_{\textit{i}}}$ in each region $i$ is computed using the ideal gas law:
\begin{equation}
T_{\rm g_{\textit{i}}} = \frac{p_{\rm g}}{\rho_{\rm g_{\textit{i}}}\mathcal{R}}, \quad \text{for } i = 1,2,3,
\end{equation}
where $\rho_{\rm g_{\textit{i}}}$ is the volume-averaged gas density in the region $i$ and $\mathcal{R}$ is the specific gas constant.
The gas density $\rho_{\rm g_{\textit{i}}}$ can be computed using the continuity equation for each region as:
\begin{equation}
\dot{m}_{\rm g_1} = -f_1, \quad \dot{m}_{\rm g_2} = f_1 - f_2, \quad \dot{m}_{\rm g_3} = f_2,
\end{equation}
where $m_{\rm g_{\textit{i}}}$ is the gas mass in the region $i$ and $f_{j}$ is the mass flux across the interface $j$:
\begin{equation}
f_j = \rho_{\rm g,\textit{j}}^{\rm uw} u_{\rm g,\textit{j}}^{\rm rel} S_j,  \quad \text{for } j = 1,2,
\end{equation}
with $\rho_{\rm g,\textit{j}}^{\rm uw}$ being the density on the upwind side of the interface $j$, $u_{\rm g,\textit{j}}^{\rm rel}$ the convective velocity (difference between real gas velocity and cell interface velocity), and $S_j$ the interface surface area.

\section{Implementation of the Francescutto \& Nabergoj model for wall-attached bubbles}\label{appE}

The instability pressure threshold $p_{\rm th, \textit{l}}$ for a shape mode with integer degree $l$, and consequently for a shape mode with order $m=l$, can be obtained under the assumption of linear oscillations by determining the stability condition of the Mathieu equation for the shape mode amplitude and reads:
\begin{equation}
p_{\rm th, \textit{l}} = \rho_{\rm l}R_0^2C_{\mathrm{th},l}\sqrt{\left(\omega_{\rm res}^2 -\omega_{\rm d}^2\right)^2+\omega_{\rm d}^2\delta^2},
\end{equation}
where $C_{\mathrm{th},l}$ is the relative bubble oscillation amplitude:
\begin{equation}
C_{\mathrm{th},l} = \sqrt\frac{\displaystyle(e-1)^2+4f}{\displaystyle\left(-\frac{3}{2}e +2f +2\left(l +\frac{1}{2}\right)\right)^2 +g^2},
\end{equation}
in which, the coefficients $e$, $f$ and $g$ are given by:
\begin{equation}
e = \frac{4(l-1)(l+1)(l+2)\sigma_{\rm lg}}{\rho_{\rm l}\omega_{\rm d}^2R_0^3},
\end{equation}
\begin{equation}
f = \left(\frac{2(l+2)(2l+1)\mu_{\rm l}} {\rho_{\rm l}\omega_{\rm d} R_0^2}\right)^2,
\end{equation}
\begin{equation}
g = \frac{6(l+2)\mu_{\rm l}}{\rho_{\rm l}\omega_{\rm d} R_0^2}.
\end{equation}
The dissipative effects are taken into account by the term $\delta$, which reads:
\begin{equation}
\delta = \frac{4\mu_{\rm e}}{\rho_{\rm l} R_0^2},
\end{equation}
where $\mu_{\rm e}$ represents the effective viscosity of the liquid, incorporating the combined contributions from viscous, thermal, and acoustic radiation losses.
Within the bubble radii range examined in this study ($\SIrange{60}{140}{\micro\meter}$) and under linear oscillation conditions, the effective viscosity is approximately 30 to 70 times greater than the contribution from liquid viscosity $\mu_{\rm l}$ alone, with these values corresponding to the smallest and largest bubbles, respectively; all values, including intermediate ones, were extracted from the theoretical plots in \cite{Chapman1971ThermalBubbles}.
Additional viscous losses may arise due to the bubble proximity to the wall and the motion of the contact line; however, these effects are difficult to quantify. Despite neglecting them, the model shows good agreement with experimental data, suggesting that their contribution is relatively minor.
Finally, $\omega_{\rm res} = 2\pi f_{\rm res}$ is the resonance angular frequency of the bubble, and $\omega_{\rm d} = 2 \pi f_{\rm d}$ is the ultrasound driving angular frequency.
The bubble resonant frequency for linear oscillations is numerically determined by solving the fully nonlinear radial dynamics model (equation~\ref{eq:RPplusImage}), incorporating the Zhou gas model, to properly account for the thermal behaviour of the bubble.
A small driving pressure ($p_{\rm a}=\SI{1}{\kilo\pascal}$) is used to ensure the bubble remains in the linear oscillation regime.
The resonant frequency is identified using a golden ratio search algorithm, which finds the driving frequency that maximises radial expansion for each bubble radius.
The resonant bubble radius corresponding to the ultrasound driving frequency employed, $f_{\rm d} = \SI{30}{\kilo\hertz}$, is $R_0 = \SI{85.8}{\micro\meter}$.

\section{Implementation of the Boundary Element Method for wall-attached bubbles}\label{appF}

Given that the Mach number $\text{M}=\dot{R}/c_{\rm l} \sim O(10^{-3})$, Reynolds number $\text{Re}=2\rho_{\rm l}\dot{R}R/\mu_{\rm l} \sim O(10^{2})$, and Bond number $\text{Bo} = (\rho_{\rm l} - \rho_{\rm g}) g R^2/\sigma \sim O(10^{-3})$ for the problem at hand, we can describe the flow velocity $\boldsymbol{u}_{\rm l}(\boldsymbol{x},t)$ and pressure $p_{\rm l}(\boldsymbol{x},t)$ within the liquid domain, denoted by~$\Omega$, using the incompressible Euler equations without considering gravitational forces:
\begin{equation}
\begin{cases}
  \boldsymbol{\nabla} \cdot {\boldsymbol{u}_{\rm l}} =0 \\
  \rho_{\rm l}\left(\displaystyle\frac{\partial{\boldsymbol{u}_{\rm l}}}{\partial t} + \left({\boldsymbol{u}_{\rm l}} \cdot {\boldsymbol{\nabla}} \right) {\boldsymbol{u}_{\rm l}}\right) = -{\boldsymbol{\nabla}} p_{\rm l}
\end{cases}
\ \text{in } \Omega.
\end{equation}
Under these conditions, the flow outside the boundary layer can be approximated as irrotational, i.e. $\boldsymbol{\nabla} \times \boldsymbol{u}_{\rm l}= 0$.
Since the domain $\Omega$ is simply connected, we can introduce a scalar potential $\phi(\boldsymbol{x},t)$ such that $\boldsymbol{\nabla}\phi = \boldsymbol{u}_{\rm l}$.
This simplifies Euler equations into the Laplace equation for the potential and the Bernoulli equation:
\begin{equation}\label{eq:PotentialFlow}
\begin{cases}
\boldsymbol{\nabla}^2\phi= 0 \\
\rho_{\rm l}\left(\displaystyle\frac{\partial \phi}{\partial t}+\frac{1}{2}\boldsymbol{\nabla}\phi \cdot \boldsymbol{\nabla}\phi\right) +p_{\rm l} = p_{\infty}+ p_{\rm d}(t)
\end{cases}
\ \text{in } \Omega.
\end{equation}

By leveraging the Green’s third identity (derived from Green’s theorem), which relates the values of a harmonic function (such as the potential $\phi$) at any point inside the domain to the values of the function and its normal derivative on the boundary, we can express the potential $\phi$ at any point $\boldsymbol{x}$ within the domain in terms of an integral over the liquid/gas interface $S$, as follows: 
\begin{equation}\label{eq:green}
c(\boldsymbol{x})\phi(\boldsymbol{x}) = \int_S \left(\phi(\boldsymbol{y}) \pder{g(\boldsymbol{x},\boldsymbol{y})}{{\boldsymbol{n}}} - \pder{\phi(\boldsymbol{y})}{{\boldsymbol{n}}}g(\boldsymbol{x},\boldsymbol{y}) \right) \mathrm{d}S(\boldsymbol{y}), \quad\forall \boldsymbol{x}\in \Omega,
\end{equation}
where $c(\boldsymbol{x})$ is the solid angle and the Green's function is given by $g(\boldsymbol{x},\boldsymbol{y}) = 1/||\boldsymbol{x} - \boldsymbol{y}||$ \citep[e.g.,][]{Steinbach:2008a}. 
Thus, the dimensionality of the problem is reduced from solving for values of the potential in the entire domain to solving only for the values along the boundary.
To solve numerically the boundary integral relation, we discretise the bubble surface using a triangular mesh, where the potential values $\phi(\boldsymbol{x}_i)$ and the normal gradients $\partial\phi(\boldsymbol{x}_i)/\partial{\boldsymbol{n}}$ at the triangle vertices $\boldsymbol{x}_i$ serve as the numerical degrees of freedom.
For computing the surface integral $\int_S\ldots\mathrm{d}S(\boldsymbol{y})$ over this triangular grid, we utilise a seven-point 2D quadrature rule for each triangle \citep[p.~295]{Radon:1948a}. To obtain the values of $\phi(\boldsymbol{y})$ and its normal derivative $\partial\phi(\boldsymbol{y})/\partial{\boldsymbol{n}}$ at surface points $\boldsymbol{y}$ within a triangle, bilinear interpolation is employed.
To address the singularities that occur in the evaluation of $g(\boldsymbol{x}_i,\boldsymbol{y})$ when the surface point $\boldsymbol{y}$ coincides with the collocation point $\boldsymbol{x}_i$, we use a polar coordinate transform on a unit triangle \citep{Kieser:1992a} in combination with a seven-point 1D Gauss--Legendre quadrature rule.
We calculate the solid angle $c(\boldsymbol{x})$ at the collocation point~$\boldsymbol{x}_i$ by applying the $4\pi$-rule, as described in \citet[eq.~(6)]{Klaseboer:2009a}.
Finally, we model the rigid wall as a mirror bubble, in accordance with the potential flow theory.

At the contact line $\boldsymbol{\Gamma}$ between the bubble surface and the solid wall, a boundary condition is imposed that allows the contact line to move along the wall plane with a velocity $w$, which is determined by the difference between the cosines of the equilibrium contact angle $\alpha_0$ and the instantaneous contact angle $\alpha(t)$.
This relationship, derived from \cite{Ren:2007a},~eq.~(28), is described by:
\begin{equation}
 w = \frac{\sigma_{\rm lg}}{\beta} \left(\cos(\alpha_0) - \cos(\alpha(t))\right),
\end{equation}
where $\beta$ is the effective friction coefficient.
This formulation leads to a condition for the normal derivative of the scalar potential $\partial \phi(\boldsymbol{x}) / \partial {\boldsymbol{n}}$ along the contact line.
To ensure numerical stability, we prevent the angle between the bubble surface and the wall from becoming too small—which corresponds to $\cos(\alpha(t))$ approaching either $1$ or $-1$—by introducing an additional nonlinear term: 
\begin{equation}
    w = \frac{\sigma_{\rm lg}}{\beta} \left(\cos(\alpha_0) - \cos(\alpha(t))\right) + c_n \frac{\cos^3(\alpha(t))}{\cos^2(\alpha(t)) - 1}.
\end{equation}
This term is designed as the simplest rational function of $\cos(\theta)$ that diverges to $+\infty$ as $\cos(\theta) \to -1$ and to $-\infty$ as $\cos(\theta) \to +1$, is antisymmetric about the origin, and has a zero slope at $\cos(\theta) = 0$. 
The ${\sigma_{\rm lg}}/{\beta}$ ratio is calibrated to 0.2 to replicate the contact line mobility observed in experiments.
The parameter $c_n$ is set to 0.01, the minimum value ensuring stable simulations.
These parameters are maintained constant in all simulations.

Ultimately, we obtain a dense linear system of equations whose unknowns are the normal gradients $\partial\phi(\boldsymbol{x})/\partial\boldsymbol{n}|_{\boldsymbol{x}=\boldsymbol{x}_i}$ at vertices away from the contact line and the potential values $\phi(\boldsymbol{x}_i)$ at vertices on the contact line. We solve this linear system using the BiCGStab solver \citep{vanderVorst:1992a}.
By combining the normal potential gradients at the interface $\partial\phi(\boldsymbol{x})/\partial\boldsymbol{n}|_{\boldsymbol{x}=\boldsymbol{x}_i}$, with the tangential gradients at the interface, which can be computed from $\phi(\boldsymbol{x}_i)$, we can numerically estimate the advection velocity at the interface nodes, $\boldsymbol{\nabla} \phi(\boldsymbol{x}_{i}) = u_{\mathrm{l}}(\boldsymbol{x}_{i})$.
Together with the contact line velocity $w$, this enables us to evolve the interface over time by advecting the nodes.
We evolve the potential on the bubble surface over time using the Bernoulli equation \citep[e.g.,][eq.~(8)]{Blake:1998a}:
\begin{equation}\label{eq:bernoulli}
\frac{\mathrm{D}\phi}{\mathrm{D}t} = \frac{1}{2}\boldsymbol{\nabla}\phi\cdot\boldsymbol{\nabla}\phi + \frac{1}{\rho_{\rm l}}\left(p_\infty +p_{\rm d}(t) - p_{\rm g,0}\left(\frac{V_0}{V(t)}\right)^\gamma + 2\sigma_{\rm lg}\kappa\right),
\end{equation}
where $D /Dt = \partial/\partial t + \boldsymbol{u}_{\rm l}\cdot \boldsymbol{\nabla} $ is the material derivative and $\kappa$ is the local bubble curvature, which is computed by using the scheme of \citet[section~3.1]{Rusinkiewicz:2004a}.
To reduce computational complexity, we assume that the gas inside the bubble undergoes an adiabatic process.
Although this model may not accurately represent the breathing mode, as demonstrated in section \ref{BrMode}, its influence on the evolution of shape modes appears to be minimal, as shown in the following.
Here, $V(t)$ represents the bubble volume at time $t$, and $V_0$ denotes the initial bubble volume.

To accurately evolve grid points and their associated potentials over time, we apply a four-stage Runge–Kutta scheme coupled with an adaptive time-stepping discretisation method.
Each simulation begins with a cut geodesic icosahedron of radius $R_0$, positioned at a height of $-R_0\cos{\alpha}$ below the wall, with all initial potential values set to zero.
The triangle mesh typically comprises several thousand elements.
After a number of time steps, the mesh quality is recalibrated using standard methods described in \cite{Botsch:2010a},~section~6.5.3.
These methods include vertex relaxation, edge flipping, and the splitting or collapsing of edges based on a specified target edge length.
In our approach, this target length is determined locally, taking into account both the local curvature and potential gradient.
To prevent abrupt changes in the target edge length between time steps, we apply a moving time-average for the target length.

\bibliographystyle{jfm}
\bibliography{main}

\end{document}